\documentclass{article}
\usepackage{arxiv}

\usepackage[utf8]{inputenc} 
\usepackage[T1]{fontenc}    
\usepackage{hyperref}       
\usepackage{url}            
\usepackage{booktabs}       
\usepackage{amsfonts}       
\usepackage{nicefrac}       
\usepackage{microtype}      
\usepackage{lipsum}
\usepackage{amssymb,amsmath,epsfig}
\newcommand{\beq}{\begin{equation}}
\newcommand{\eeq}{\end{equation}}
\newcommand{\bea}{\begin{eqnarray}}
\newcommand{\eea}{\end{eqnarray}}

\begin{document}
\title{Shadow cast by a rotating charged black hole in quintessential dark energy}

\author{{Saeed Ullah Khan\thanks{saeedkhan.u@gmail.com}} \, and \, {Jingli Ren\thanks{Corresponding author: renjl@zzu.edu.cn}}\vspace{0.2cm} \\\vspace{0.08cm}
Henan Academy of Big Data/School of Mathematics and Statistics, Zhengzhou University,\\
Zhengzhou 450001, China.}
\date{}
\maketitle
\begin{abstract}
The existence of quintessential dark energy around a black hole has considerable consequences on its spacetime geometry. Hence, in this article, we explore its effect on horizons and the silhouette generated by a Kerr-Newman black hole in quintessential dark energy. Moreover, to analyse the deflection angle of light, we utilize the Gauss-Bonnet theorem. The obtained result demonstrates that, due to the dragging effect, the black hole spin elongates its shadow in the direction of the rotational axis, while increases the deflection angle. On the other hand, the black hole charge diminishing its shadow, as well as the angle of light's deflection. Besides, both spin and charge significantly increase the distortion effect in the black hole's shadow. The quintessence parameter $\gamma$, increases the shadow radius, while decreases the distortion effect at higher values of charge and spin parameters.
\end{abstract}

\section{Introduction}
In recent years, inspecting the invisible aspects of our Cosmos got a considerable interest of many researchers. Various cosmic data-sets specified that dark energy (DE) is a crucial source for the accelerated expansions of our universe \cite{Peebles}. About $70\%$ of our observable universe is consists of DE \cite{Caldwell1}, which could be explained with the help of a repellent cosmological constant $\Lambda>0$ or by a quintessential field \cite{Ostriker,Caldwell2,Faraoni,Arm,Riess,Ada}. The parameter $\Lambda$ can be assumed as homogeneous, while everywhere in space, its value remains the same ($\Lambda\approx 1.3\times 10^{-56} cm^{-2}$) \cite{Stuch}. Apart from cosmological constant, another essential model of the DE is known as the quintessence \cite{Copeland}. Spacetime geometry has greatly influenced by the contents of DE, i,e., a quintessence field and/or a cosmological constant of the black hole (BH). In the presence of a cosmological constant, the asymptotic structure of a BH changes to the asymptotic de Sitter spacetime \cite{Stuchlik1,Xu}.

The idea of BHs shadow was initiated by Bardeen \cite{Bardeen} with the conclusion that a BH has a radius of $r_{shadow}=5.2M$, over a background light source (visible to an external observer). As described by many researchers, the size of shadows cast by spinning BHs is almost the same as well \cite{Takahashi,Hioki,Johannsen,Amarilla,Moffat,Sharif1}. Theoretical investigation of the shadow cast by the BH horizons can be studied as the existence of a photon sphere and null geodesics. To an external observer, BH appears to be a dark disc termed as the shadow of a BH. In the case of non-rotating BHs, the shadow appeared to be a circular disc, whereas, in rotating BHs the shadow appeared to be flattened on one side, rather than a circular disc \cite{Perlick}. In recent years, many researchers have been motivated by the astrophysical advances to investigate shadows cast by BHs \cite{Vries,Abdu1,Haroon,Konoplya}. Researchers accept that very soon, direct examinations of the BHs could be possible \cite{Cash,Doeleman,Falcke,Bambi}. Henceforth, studying a BH's shadow will be a fruitful way for a better understanding of astrophysical BHs, as well as for the comparison of general relativity to those of modern theories \cite{Eiroa}. 
To have information on BHs, the approaches of null geodesics and gravitational lensing are of compelling interest \cite{Ovgun1,Shahzadi,Khan}. Recently, scientists have finally succeeded in obtaining the first-
ever image of a supermassive BH, at the centre of the M87 galaxy \cite{ETH,Akiyama1,Akiyama2}. Among other evidence, this is the strongest ever assurance to the existence of supermassive BHs, which opens up new windows onto the investigation of BHs.

Gravitational lensing (GL), can be defined as the bending of light due to gravity. They can be used to examine the distribution of dark matter in our Cosmos, as well as to explain the far-away galaxies. Using the strong GL, we can detect the location, magnification and 
time delays of shadows by BHs, whereas, in weak GL the consequence is much weaker yet could still be analysed statistically \cite{Hu}. Moreover, the weak, as well as strong GL by wormholes and BHs, can be found in \cite{Virbhadra,Stefanov,Ono1}. The assumption that massive particle bends light rays while passing through it is remarkable in general relativity. The applications of GL consist of exploring the deflection angle, as well as detecting BHs in our Cosmos. Gibbons and Werner \cite{Gibbons} by making use of the Gauss-Bonnet theorem (GBT), introduced a new approach to to acquired the light's deflection angle. In this approach, bending of light can be viewed as a global topological effect, rather than corresponding to a region with a radius compared to the impact parameter. Besides BHs, the Gibbon-Warner approach for the deflection angle of light ray has also studied for both of the asymptotically flat and non-flat spacetimes and wormholes \cite{Jusufi,Javed1,Li}.

In current work, our primary focus is on a non-singular domain exterior to a light ray. In the case of an asymptotically flat spacetime geometry, the deflection angle $\hat{\Theta}$ could be calculated as \cite{Gibbons}
\begin{equation}\nonumber
\hat{\Theta} =-\int \int_{_R^\infty\square _S^\infty} K dS,
\end{equation}
where $K$ and $dS$, respectively denote the Gaussian optical curvature and the surface element of optical geometry. It should be noted that the above expression of $\hat{\Theta}$, can only be satisfied in asymptotically flat spacetimes, where one can only consider a finite distance correction for the non-asymptotically flat spacetime geometries.

In this article, besides charge and the rotational parameter $a$, our main goal is to understand the effects of quintessential DE on BH shadow. The following section will provide a brief review of the Kerr-Newman (KN) BH in quintessential DE and a cosmological constant (KNdSQ BH) and the angle of light's deflection. In section \ref{sec:geodesics}, we will investigate the null geodesics in detail. The key objective of section \ref{sec:shadow} is to explore the images generated by a KNdSQ BH. Finally, the last section will provide discussions and a thorough summary of our obtained results.
\section{Spacetime metric of the KNdSQ BH}\label{metric}
Motivated by previous work, this section aimed to investigate the KNdSQ BH. The KNdSQ BH is the solution of Einstein-Maxwell equation and in Boyer-Lindquist coordinates takes the form \cite{Xu}
\begin{eqnarray}\label{e1}
ds^2=\frac{\Delta _\theta \sin^2\theta}{\rho^2}{\left(a\frac{dt}{\Sigma}-\left(a^2+r^2\right)\frac{{d\phi}}{\Sigma }\right)^2}+\frac{\rho^2}{\Delta_r}{dr}^2 +\frac{\rho^2}{\Delta_\theta}{d\theta}^2  - \frac{\Delta _r}{\rho ^2}{ \left(\frac{dt}{\Sigma}-a\sin^2\theta \frac{d\phi}{\Sigma }\right)^2},
\end{eqnarray}
with
\begin{eqnarray}\nonumber
\Delta_r &= &r^2-2Mr+a^2+Q^2-\gamma r^{1-3\omega}-\frac{\Lambda}{3}(r^2+a^2)r^2, \\\nonumber
\rho^2 &=&r^2+a^2 \cos^{2}\theta, \quad \Delta_\theta = 1+ \frac{a^2}{3}\Lambda \cos^2\theta, \quad \Sigma = 1+\frac{a^2}{3}\Lambda.
\end{eqnarray}
In the above model, $M$, $a$ and $\Lambda$, respectively represent mass, spin and cosmological constant of the BH. The parameter $Q$ denotes BH charge, while $\gamma$ is the intensity of the quintessence field of the BH.
Metric \eqref{e1}, reduces to the KN spacetime by substituting $\gamma=\Lambda=0$; to the Kerr BH if $\gamma=\Lambda=Q=0$; to the RN BH if $\gamma=\Lambda=a=0$; and to the Schwarzschild BH by letting $\gamma=\Lambda=a=Q=0$.
The horizons of metric \eqref{e1}, are the null hypersurfaces caused by null geodesics of the BH and can be obtained by solving $\Delta_r=0$, as
\beq\label{EH}
r^2-2Mr+a^2+Q^2-\gamma r^{1-3\omega}-\frac{\Lambda}{3}(r^2+a^2)r^2=0.
\eeq
As previously discussed by Xu and Wang, the parameters $\gamma$, $a$, $Q$, $\Lambda$ and $\omega$ defines the number of horizons \cite{Xu}. Moreover, its horizons are different from that of Kerr BH. The horizons of KN BH can be recovered by substituting $\gamma=\Lambda=0$. The horizons of Kerr and Schwarzschild can be obtained by putting $\gamma=\Lambda=Q=0$ and $\gamma=\Lambda=a=Q=0$, respectively.

\begin{figure*}
\begin{minipage}[b]{0.58\textwidth} \hspace{0.1cm}
        \includegraphics[width=0.78\textwidth]{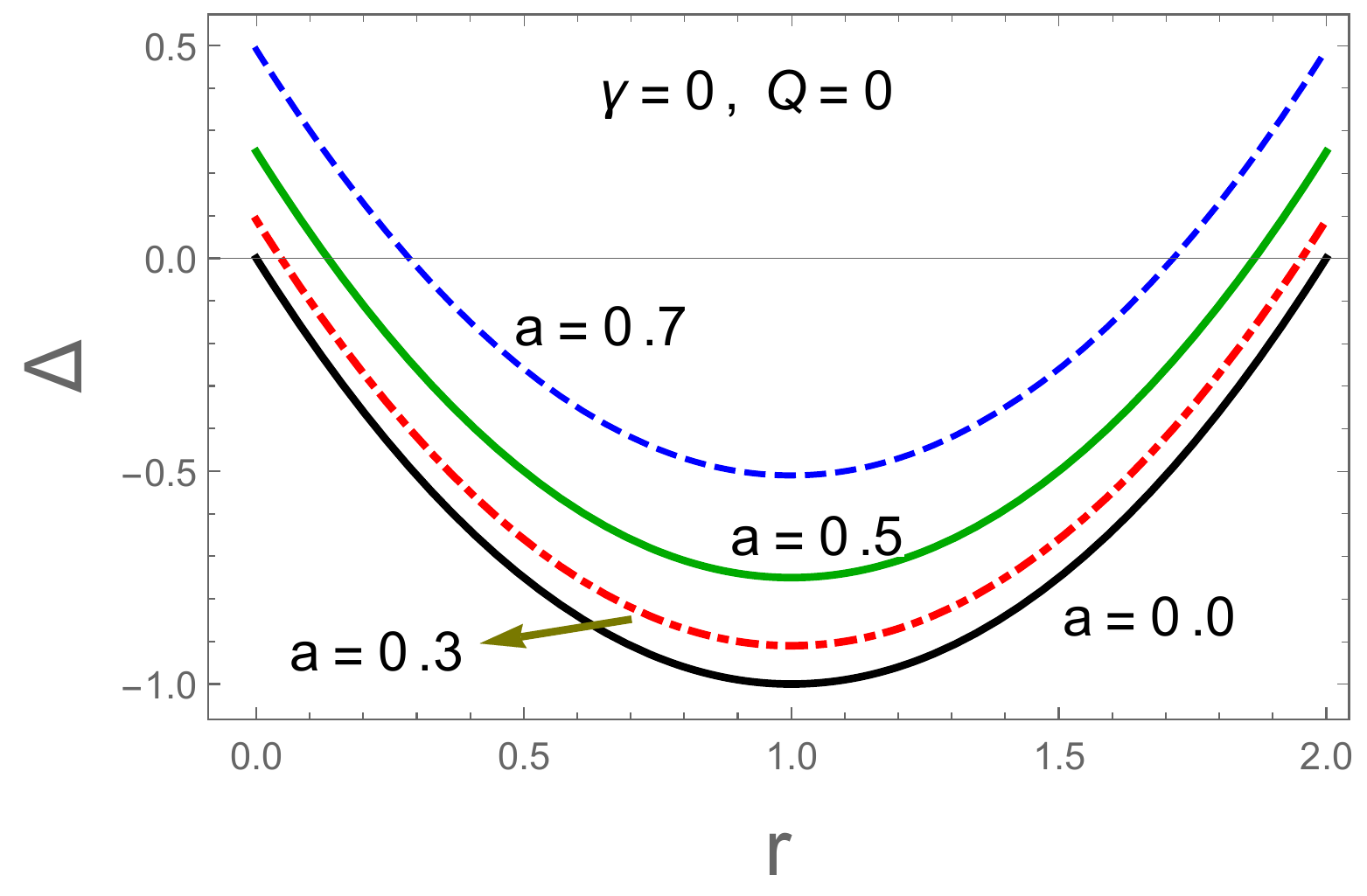}
    \end{minipage}
    \vspace{0.2cm}
        \begin{minipage}[b]{0.58\textwidth} \hspace{-1.2cm}
       \includegraphics[width=.78\textwidth]{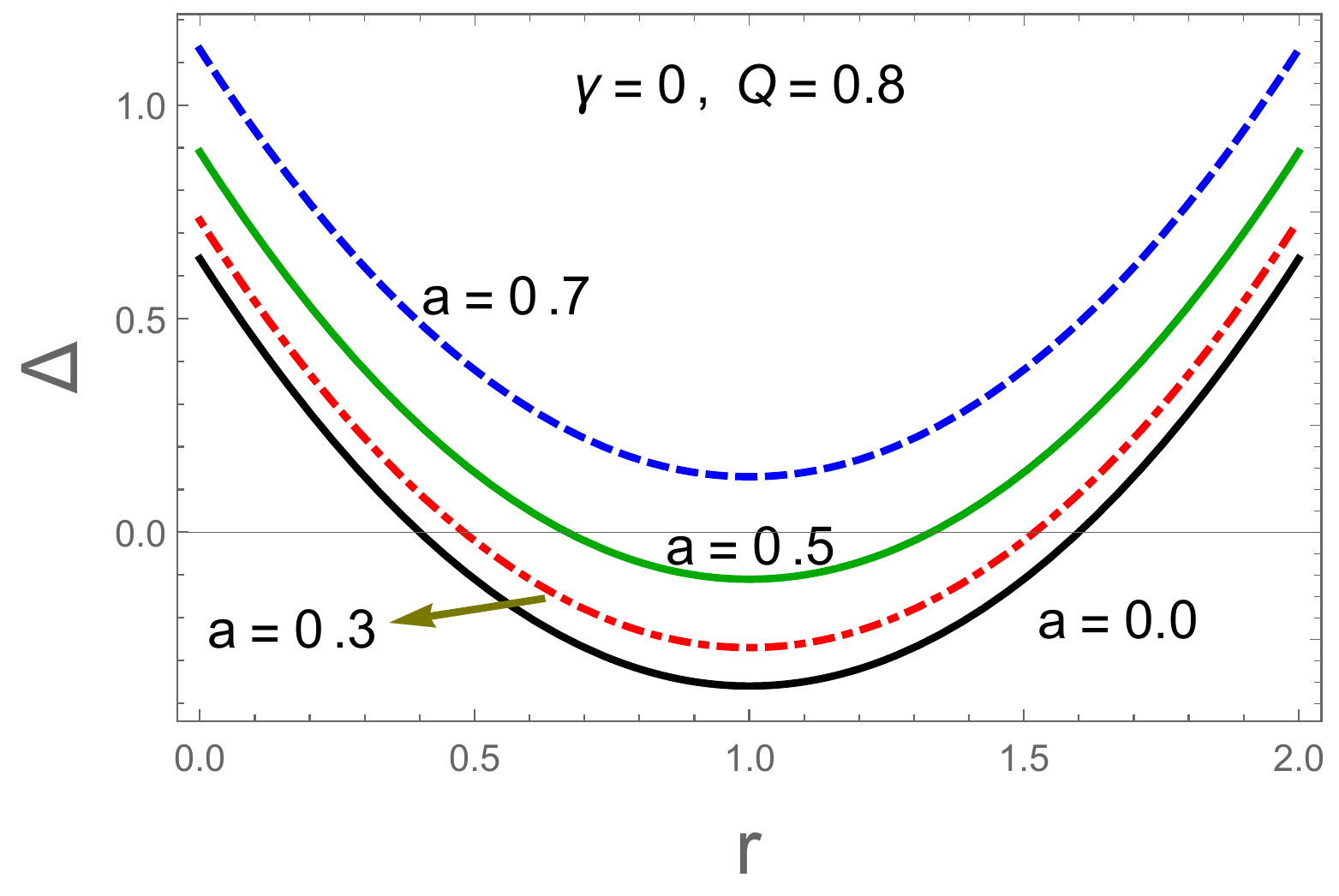}
    \end{minipage}
\begin{minipage}[b]{0.58\textwidth} \hspace{0.1cm}
        \includegraphics[width=0.78\textwidth]{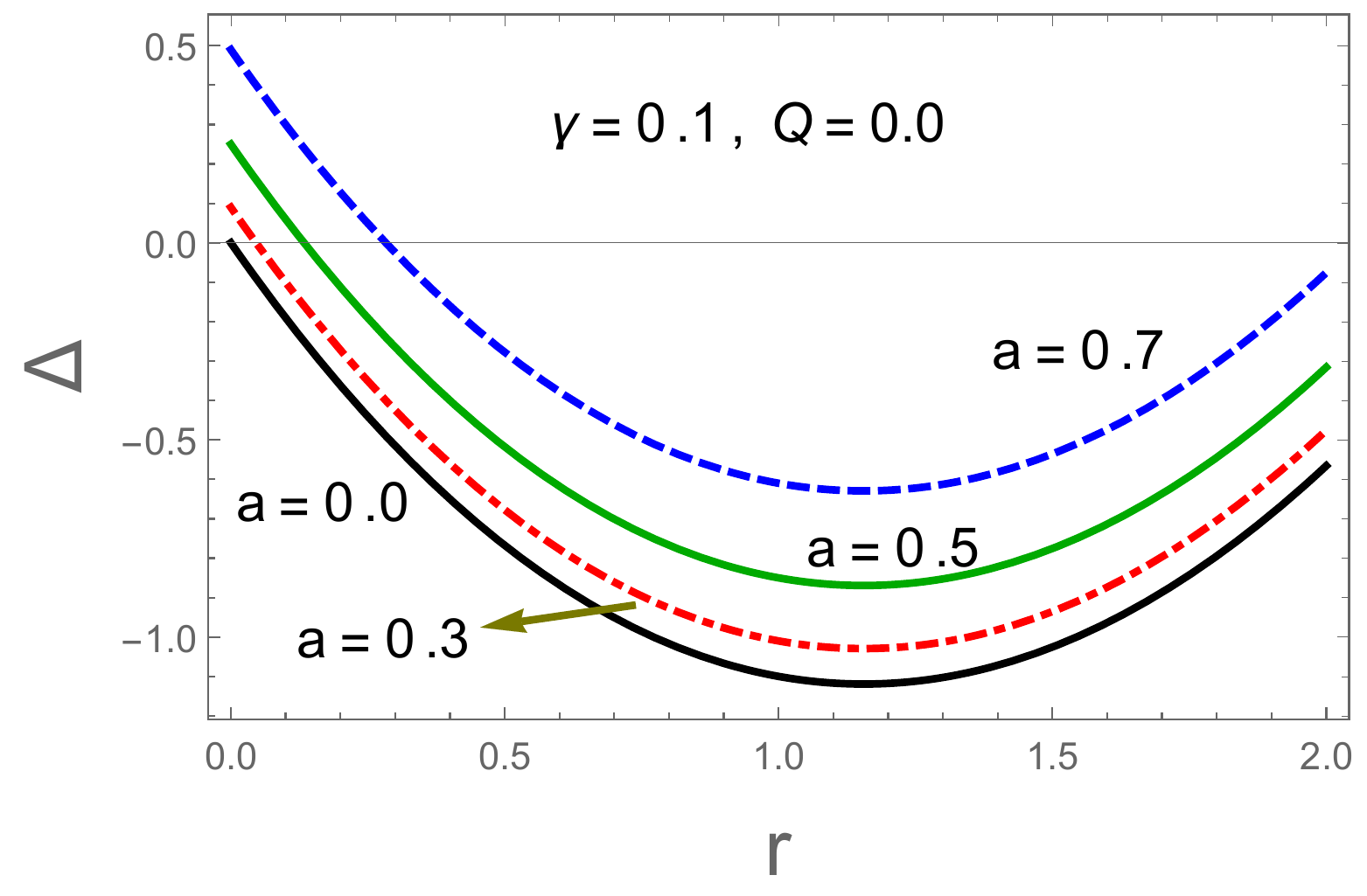}
    \end{minipage}
    \vspace{0.2cm}
        \begin{minipage}[b]{0.58\textwidth} \hspace{-1.2cm}
       \includegraphics[width=.78\textwidth]{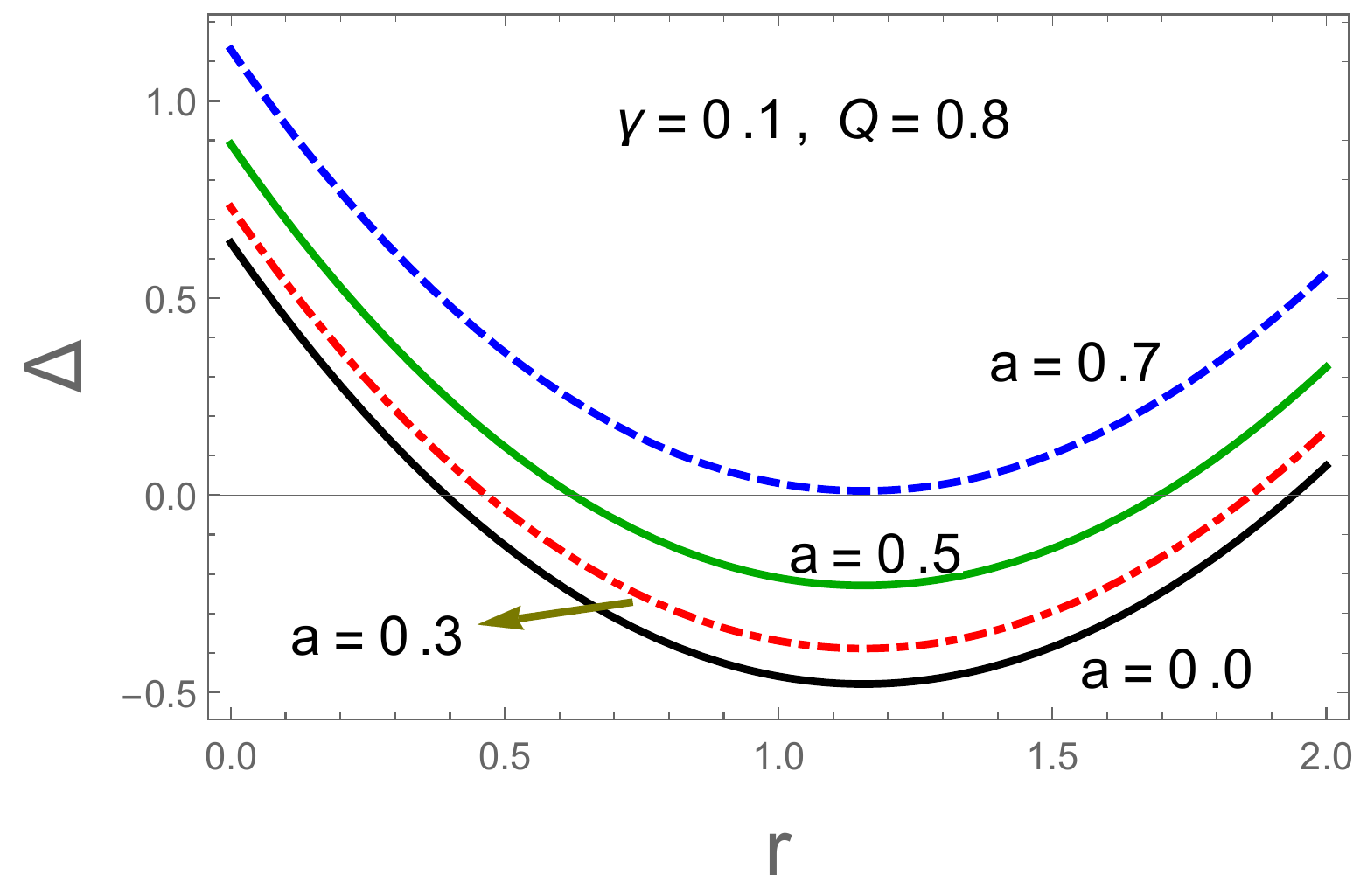}
    \end{minipage}
    \begin{minipage}[b]{0.58\textwidth} \hspace{0.1cm}
        \includegraphics[width=0.78\textwidth]{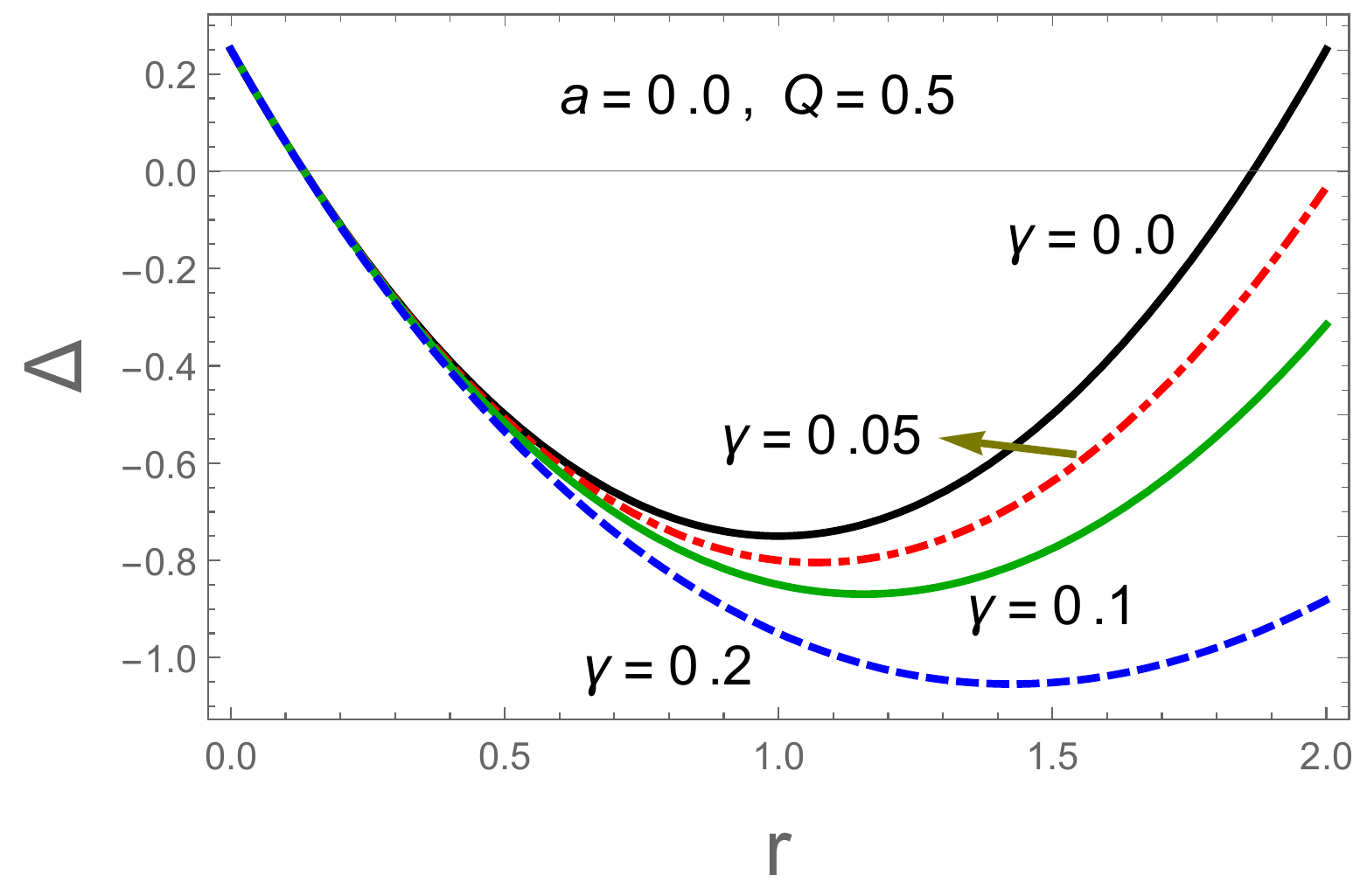}
    \end{minipage}
        \begin{minipage}[b]{0.58\textwidth} \hspace{-1.2cm}
       \includegraphics[width=.78\textwidth]{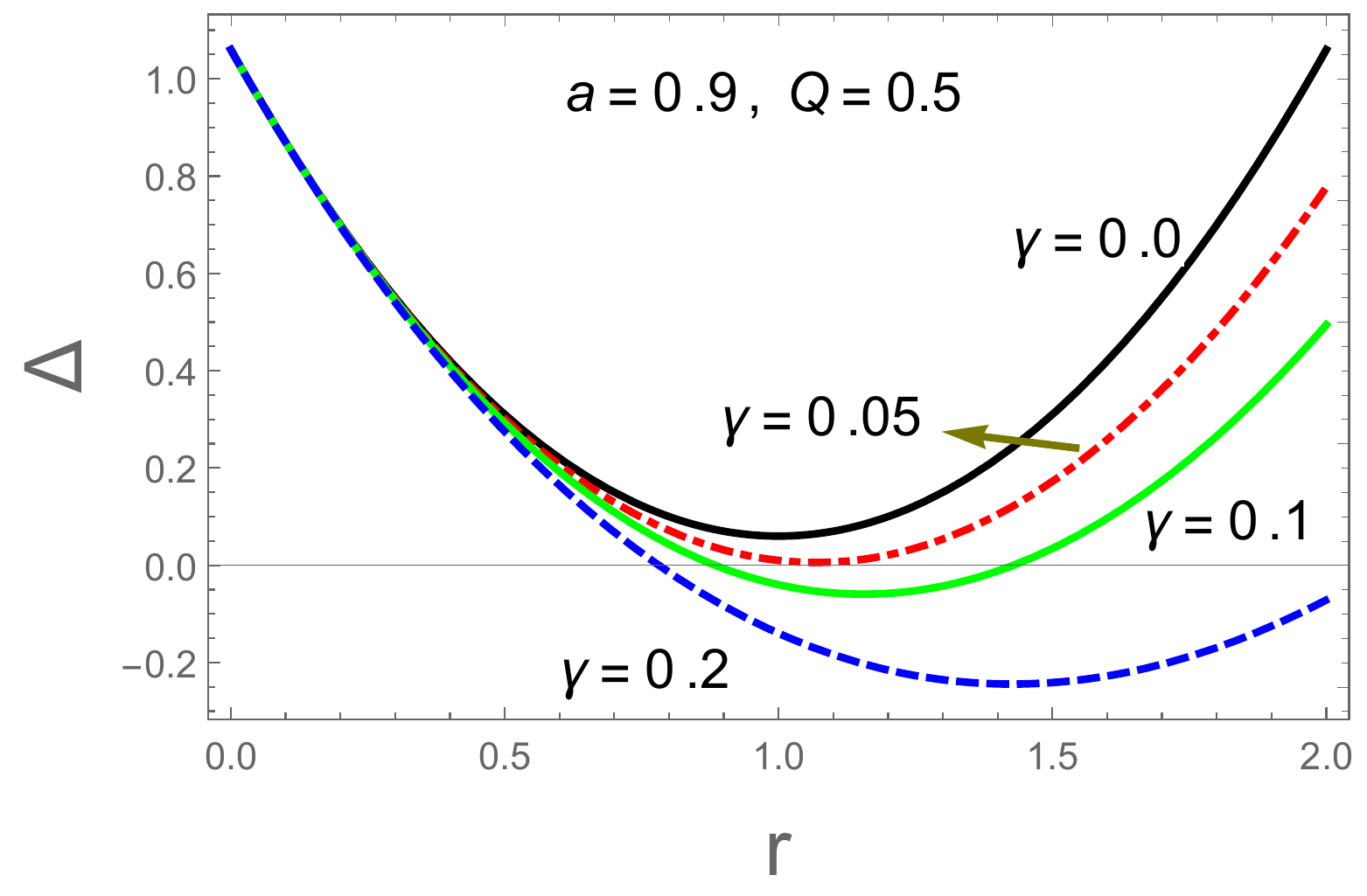}
    \end{minipage}
\caption{The horizons structure at different values of the BH parameters.}\label{horizons}
\end{figure*}

\begin{figure*}
\begin{minipage}[b]{0.58\textwidth} \hspace{0.1cm}
        \includegraphics[width=0.78\textwidth]{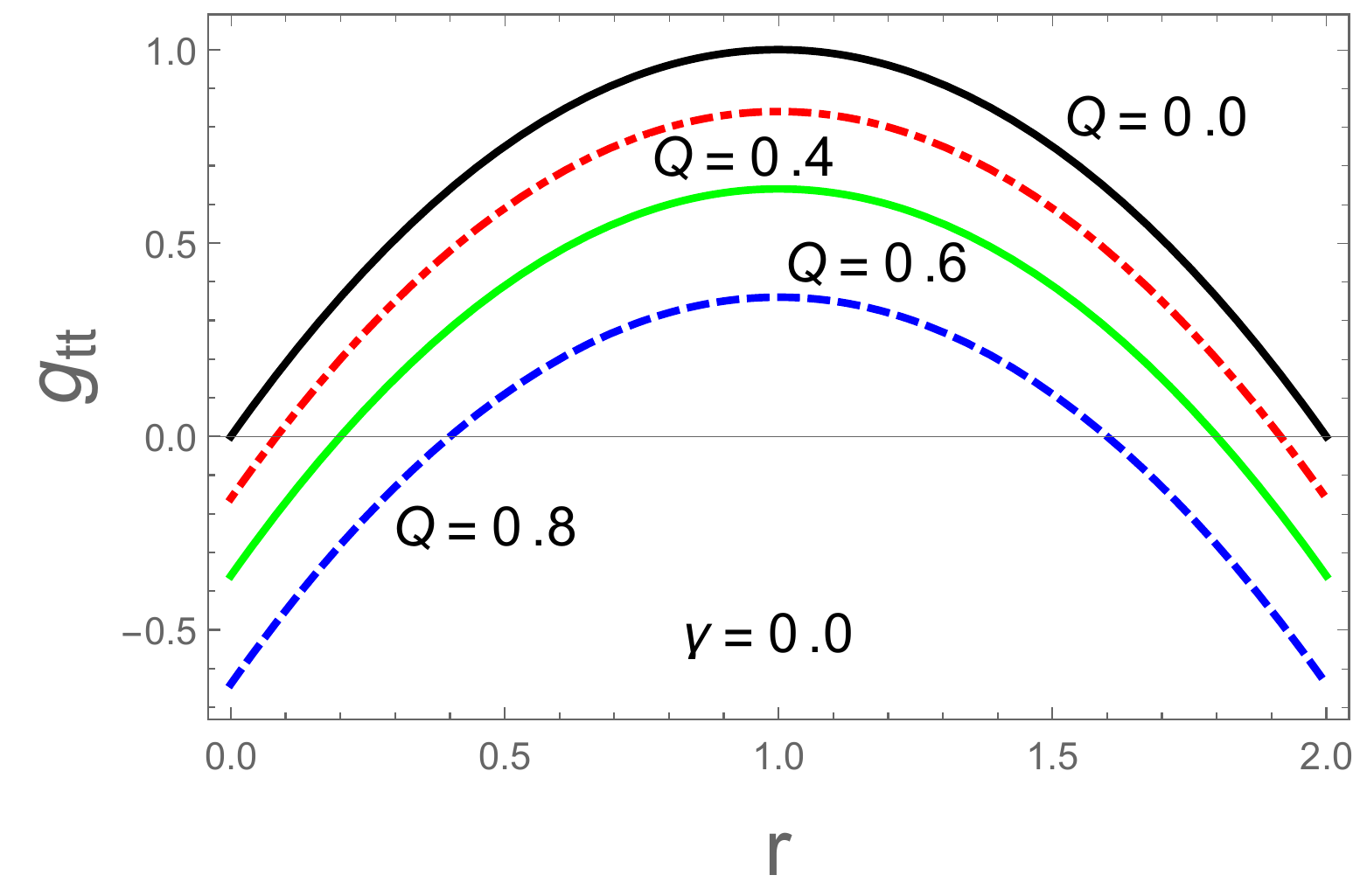}
    \end{minipage}
    \vspace{0.2cm}
        \begin{minipage}[b]{0.58\textwidth} \hspace{-1.2cm}
       \includegraphics[width=.78\textwidth]{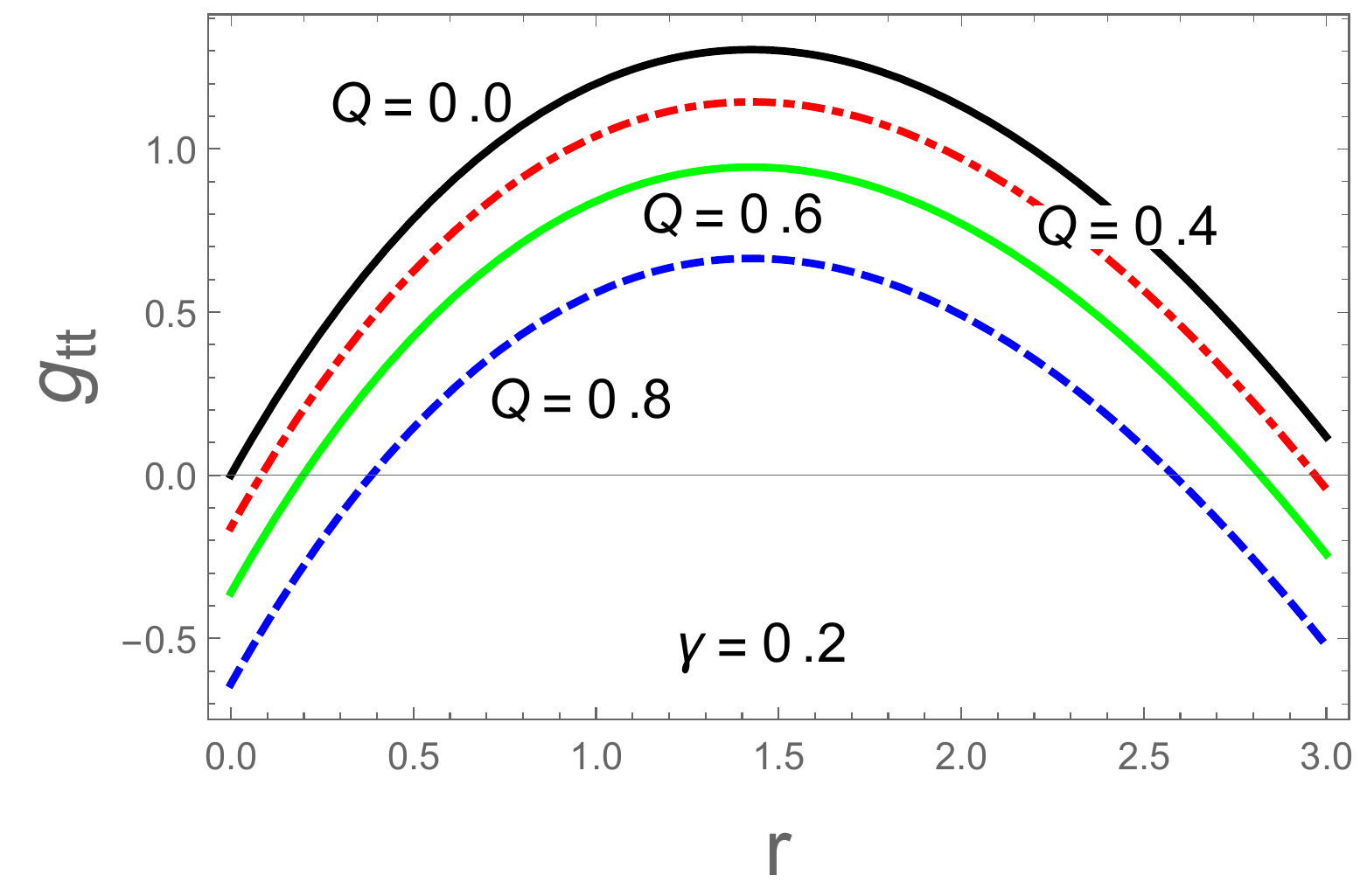}
    \end{minipage}
\begin{minipage}[b]{0.58\textwidth} \hspace{0.1cm}
        \includegraphics[width=0.78\textwidth]{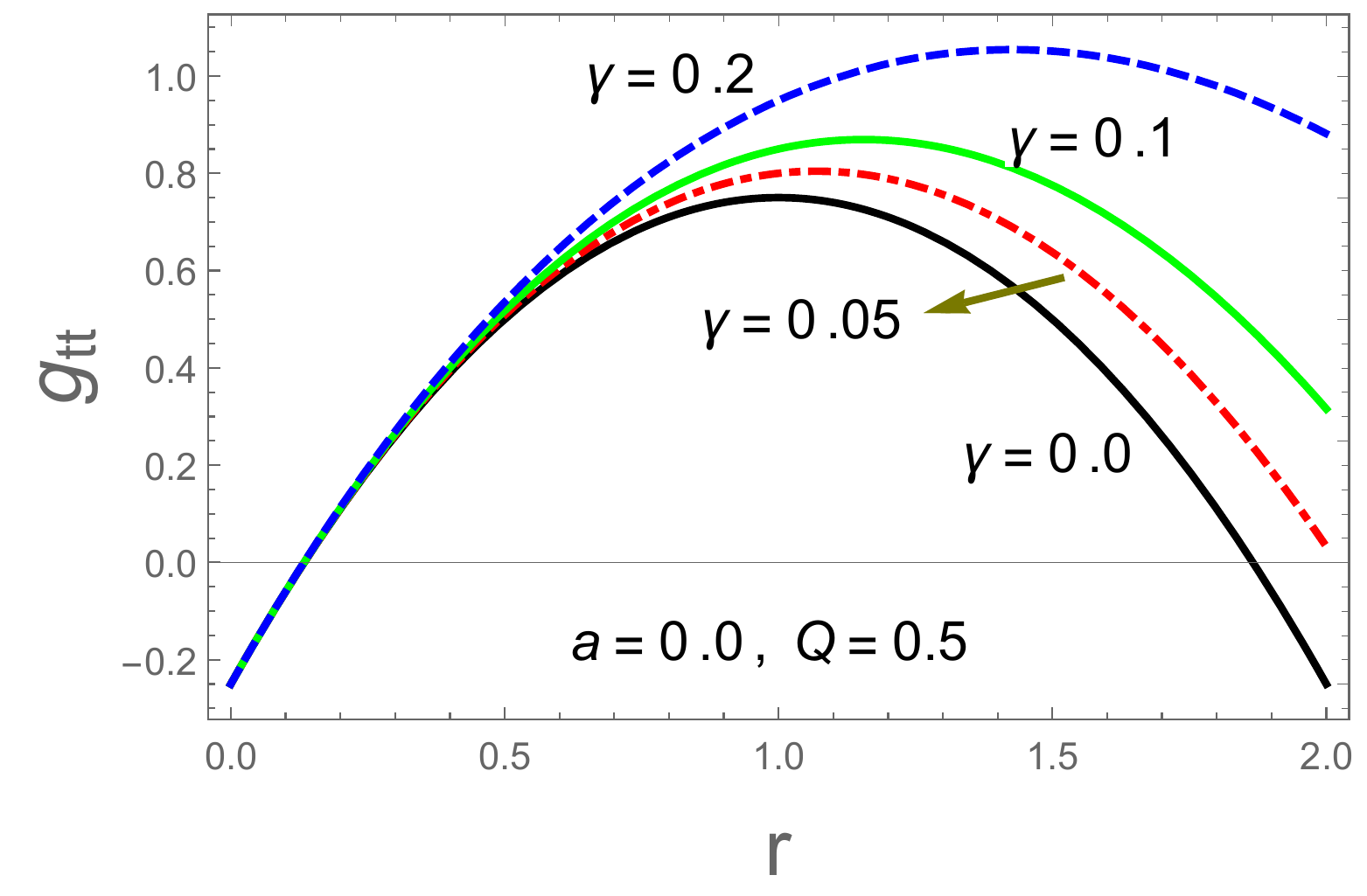}
    \end{minipage}
        \begin{minipage}[b]{0.58\textwidth} \hspace{-1.2cm}
       \includegraphics[width=.78\textwidth]{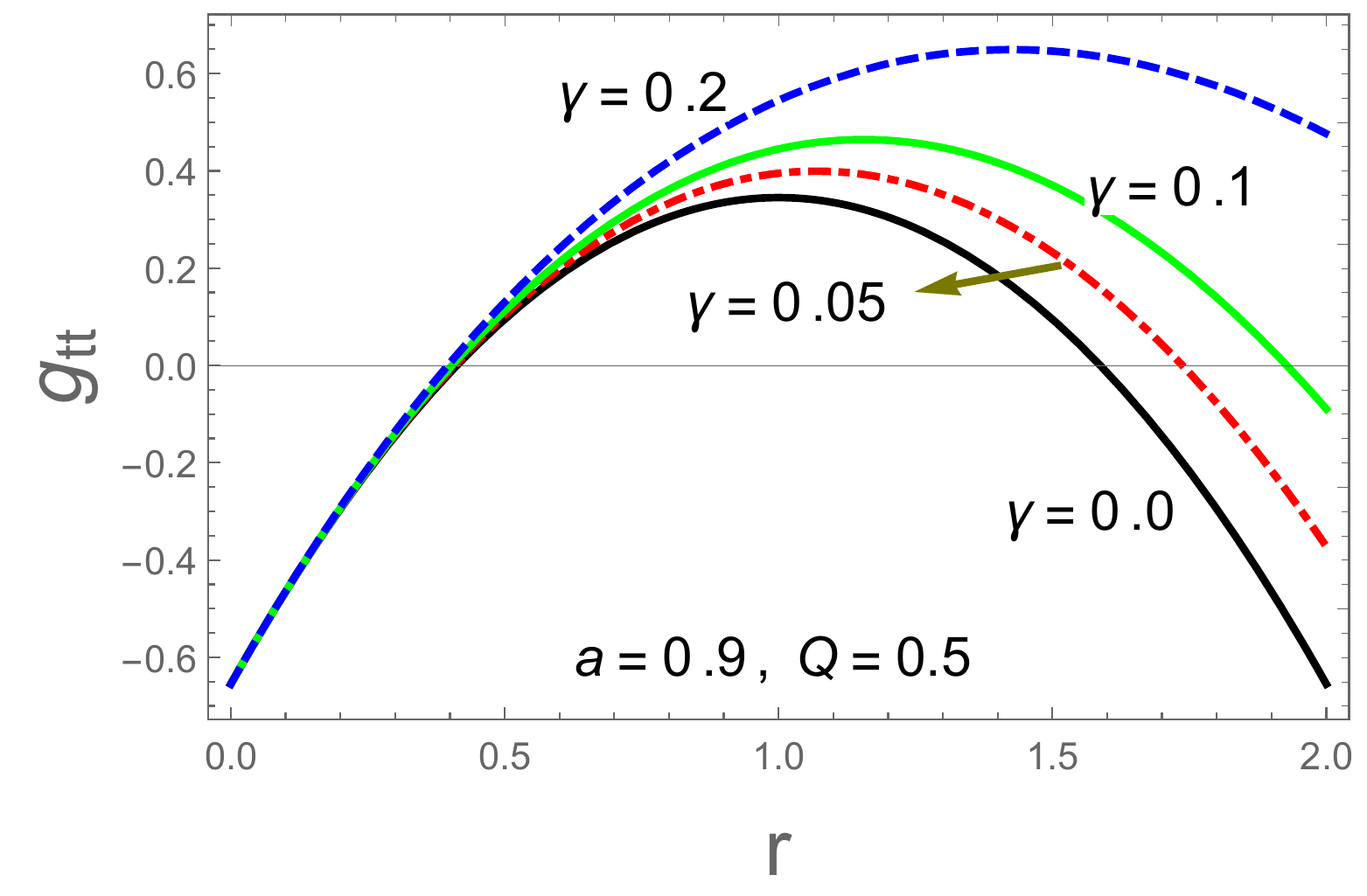}
    \end{minipage}   
\caption{The upper row is plotted for $\theta = \pi/2$ and $a=0$, while the lower row is plotted for $\theta = \pi/4$.}\label{SL}
\end{figure*}

The horizons structure of a KNdSQ spacetime has shown in Fig. \ref{horizons}. The first row of Fig. \ref{horizons}, depicts that both BH rotation and charge increase the value of $\Delta_r$, whereas $\gamma$ elongates the geometrical structure of its horizons. From the third row, we observed that the quintessence field $\gamma$ contributes to the BH horizons. Our investigation shows that as compared to the KN BH, Kerr has greater horizons. Moreover, the rotating case has smaller horizons as compared to the non-rotating case. 
The ergosphere can be calculated by letting $g_{tt} = 0$, or
\beq\label{SLS}
a^2\sin^2{\theta \Delta_{\theta}}-\Delta_r=0,
\eeq
which is the boundary of the static limit surface (SLS). Besides, other parameters the SLS also depends on $\theta$. On substituting $\gamma=\Lambda=0$, the SLS reduces to the case of KN BH and admit the solution
\beq
r_{SLS\pm}=M\pm \sqrt{M^2-Q^2-a^2\cos^\theta}.
\eeq
Figure \ref{SL} reflects the numerical investigation of Eq. \eqref{SLS}, which describes the behaviour of $g_{tt}$, along with the radial distance $r$. We noted that both BH charge and spin reduces its SLS, while the quintessence parameter $\gamma$ elongates and increase the SLS.
\subsection{Deflection of light}
\label{Subsec:deflection}
To analyse the angle of light’s deflection, we make use of the GBT. For this purpose, we are using the null condition $ds^2=0$ on the equatorial plane $(\theta=\pi/2)$. For simplicity, we substitute $\Lambda=0$ and by following \cite{Ishihara1} define
\begin{equation}\label{DA1}
dt^2=\sqrt{\gamma_{ij}dx^i dx^j}+\beta_idx^i.
\end{equation}
Here $\gamma_{ij}$ goes as $i,j=1,2,3 $, and the corresponding component of optical metric can be obtained as
\begin{eqnarray}\nonumber
d\ell^2 = \frac{\rho^4}{\Delta_r\left({\Delta_r}-a^2 \right)}dr^2 + \left(\frac{(a^2+r^2)-a^2\Delta_r}{\Delta_r-a^2} +\frac{a^2(\Delta_r+(a^2+r^2))^2}{\Delta_r-a^2} \right)d\phi^2,
\end{eqnarray}
with
\begin{equation}\nonumber
\beta_idx^i=\frac{a^2(\Delta_r-(a^2+r^2))}{\Delta_r-a^2}d\phi.
\end{equation}
Let $\Psi_S$ and $\Psi_R$, respectively be the angles of source and receiver, while $\Phi_{RS}$ be the coordinate angle defined by $\Phi_{RS}=\Phi_R-\Phi_S$. Henceforth, the deflection angle of light can be defined as \cite{Ishihara1}
\begin{equation}\label{DA2}
\hat{\Theta} \equiv \Psi_R-\Psi_S+\Phi_{RS}.
\end{equation}
To find out the point of source and receiver, where the light rays endpoints exist in the Euclidean space. Let $X$ and $\partial X_i$, respectively denote a $2D$ orientable surface and the curves of its orientable boundaries, whereas jump angles between the curves are expressed by $\theta_i$. Henceforth, in terms of the GBT, the deflection angle can be defined as \cite{Ishihara2}
\begin{equation}\label{DA3}
\int \int_{X}K dS+\sum_{i=1}^N \int_{\partial X_i}k_g d\ell+\sum_{i=1}^N\theta_i=2\pi.
\end{equation}
Here $K$, $k_g$, are respectively the Gaussian and geodesics curvatures of $X$ and $\partial X_i$. The line element of the boundary is denoted by $\ell$, while its sign is chosen in correspondence with the surface orientations. Following the GBT, we can define (see for detail \cite{Ishihara2})
\begin{eqnarray}\label{DA4}
\hat{\Theta}= \Psi_R-\Psi_S+\Phi_{RS}=-\int \int_{_R^\infty\square _S^\infty}K dS.
\end{eqnarray}
In the above equation, ${_R^\infty\square _S^\infty}$ denotes a quadrilateral and after simplification, the deflection angle of KNdSQ BH simplifies to \cite{Javed}
\begin{equation}\label{Da5}
\hat{\Theta} \equiv \frac{4M}{b}+\frac{2\gamma}{b}-\frac{3\gamma\omega}{b}-\frac{3Q^2}{4b^2}\pm\frac{4aM}{b^2}.
\end{equation}
Here the $\pm$ sign indicates the indirect and direct case of photon orbits. The deflection angle reduces to the case of KN BH by letting $\gamma=0$; to the case of Kerr BH if $\gamma=Q=0$ \cite{Kerr}; and to the case of well known Schwarzschild BH if $a=\gamma=Q=0$ \cite{Schwarz}.
\section{Null geodesics}
\label{sec:geodesics}
Generally, the shadow of a BH can be defined as the boundary of photon capture and scattering orbits \cite{Chandrasekhar}.  In our case, we consider photon orbits around a KNdSQ BH, with the condition of $p^{\mu}p_{\mu}=-m^2$. Henceforth, by making use of the Lagrangian equation
\begin{equation}\label{BS7}
\mathcal{L} = \frac{1}{2}g_{\mu\nu}\dot{x}^{\mu}\dot{x}^{\nu},
\end{equation}
here dot means derivative with respect $\tau$ ($\tau$ is the proper time) and $\dot{x}^{\mu}={dx^{\mu}}/{d\tau}$ where $x^\mu$ represent the corresponding four velocities of the particle. BH symmetry allows us to define the conserved energy and angular momentum as,
\begin{eqnarray}\label{BS8}
\frac{\partial\mathcal{L}}{\partial\dot{t}}&=&p_{t}=g_{tt}\dot{t}+g_{t\phi} \dot{\phi}=\mathcal{E},\\ \label{BS9}
\frac{\partial\mathcal{L}}{\partial\dot{\phi}}&=&p_{\phi}=g_{t\phi}\dot{t}+g_{\phi\phi}\dot{\phi}=-L_z,
\end{eqnarray}
On solving the above equations for $\dot{t}$ and $\dot{\phi}$, leads us to the following two equations 
\begin{eqnarray}\label{BS10}
\rho^2\Sigma^2\frac{dt}{d\tau}&=& a(L_z-a \mathcal{E}\sin^2\theta)+\frac{(a^2+r^2)}{\Delta_r}\left((a^2+r^2)\mathcal{E}-aL_z \right),\\\label{BS11}
\rho^2\Sigma^2\frac{d\phi}{d\tau}&=&  \left(L_z\csc^2\theta-a \mathcal{E} \right)+\frac{a}{\Delta_r}\left((a^2+r^2)\mathcal{E}-aL_z \right).
\end{eqnarray}
For obtaining the other two geodesic equations, we follow the Hamiltonian-Jacobi approach and obtain the equations of geodesics as   
\begin{equation}\label{BS1}
\frac{1}{2}g_{\mu \nu}\frac{\partial S}{\partial x^{\mu}}\frac{\partial S}{\partial x^{\nu}} +\frac{\partial S}{\partial \tau}=0,
\end{equation}
and Carter's separability prescription \cite{Carter}, we select the corresponding action as
\begin{equation}\label{BS2}
S=\frac{1}{2}m_0^2\tau-\mathcal{E}t+L_z\phi+S_r(r)+S_\theta(\theta).
\end{equation}
Here $m_0$ represents photon mass and by solving Eq. \eqref{BS2} for $S_r$ and $S_\theta$, we get
\begin{eqnarray}\label{BS3}
\rho^2\Sigma^2\frac{\partial S_r}{\partial r}&=&\sigma_r \sqrt{\mathcal{R}(r)},\\\label{BS4}
\rho^2\Sigma^2\frac{\partial S_\theta}{\partial \theta}&=& \sigma_\theta \sqrt{\Theta(\theta)},
\end{eqnarray}
where $\sigma_r=\sigma_\theta=\pm$, with 
\begin{eqnarray}\label{BS5}
\mathcal{R}(r)&=&\left((a^2+r^2)\mathcal{E}-a L_z\right)^2-\Delta_r \left((L_z-a\mathcal{E})^2+\mathcal{O} \right),\\\label{BS6}
\Theta(\theta)&=&\mathcal{O} +\left(a^2 \mathcal{E}^2-L_z^2 \csc^2\theta \right)\cos^2\theta.
\end{eqnarray}
In the above equations $\mathcal{O}=\mathcal{K}-(a\mathcal{E}-L_z)^2$, represents Carter's constant in which $\mathcal{K}$ is the constant of motion. Null geodesics near a KNdSQ BH is governed by the above four Eqs. \eqref{BS10}, \eqref{BS11}, \eqref{BS3} and \eqref{BS4}. They particularly depend on the two impact parameters defined in terms of $\mathcal{E}$ and $L_z$ as $\xi=L_z/\mathcal{E} $ and $\eta=\mathcal{O}/\mathcal{E}^2$ \cite{Chandrasekhar}. For a photon, there exist three different types of geodesics known as scattering, spherical and plumbing orbits; and for the case of photons the radial Eq. \eqref{BS5}, can be rewritten in terms of $\xi$ and $\eta$ as
\begin{equation}\label{BS12}
\mathcal{R}(r)=\frac{1}{\mathcal{E}^2}\left[ \left((a^2+r^2)- a\xi \right)^2 -\Delta_r\left((a-\xi)^2+\eta \right) \right].
\end{equation}
Using the above Eq. \eqref{BS12}, the required effective potential on a photon can be acquired. Henceforth, the critical and unstable circular orbits can be obtained with the help of maximum effective potential, satisfying the constraints of
\begin{eqnarray}\label{BS13}
\mathcal{R}(r)=\frac{\partial \mathcal{R}(r)}{\partial r}\mid_{r=r_0}=0.
\end{eqnarray}
Here $r_0$ denotes the radius of the circular unstable null orbit. We assumed that both photon and observer are located at infinity. The value of celestial coordinates $\xi$ and $\eta$, can be calculated from Eq. \eqref{BS13}, as
\begin{eqnarray}\nonumber\label{BS14}
\xi&=&\frac{1}{2ar_0^{3\omega}\left(r_0\left(a^2\Lambda+2\Lambda r_0^2-3\right)+3M\right)+3a\gamma(1-3\omega)} [3\gamma(a^2(1-3\omega)\\\nonumber &&-3r_0^2(\omega +1))+2 r_0^{3 \omega}(a^4\Lambda r_0+a^2(3(M+r_0)+\Lambda r_0^3)+3r_0(r_0(r_0\\&& -3 M)+2 Q^2))], \\\nonumber\label{BS15}
\eta&=&\frac{-1}{\left(2ar_0^{3\omega}\left(r_0\left(a^2\Lambda+2\Lambda r_0^2-3\right)+3M\right)+3a\gamma(1-3\omega)\right)^2} [r_0^2(4a^4 \Lambda^2 r_0^{6\omega+4}\\\nonumber && +12a^2r_0^{3\omega}(2r_0^{3\omega}\left(\Lambda r_0^3(3M+r_0)-6Mr_0+Q^2\left(6-2 \Lambda r_0^2\right)\right)+3\gamma r_0(\Lambda r_0^2(1 \\&& +\omega)-6\omega-2))+9 \left(2r_0^{3\omega}\left(r_0(r_0-3M)+2Q^2\right)-3\gamma r_0(\omega +1)\right)^2)].\quad
\end{eqnarray} 
It should be noted that the above impact parameters are essential, as they determine the boundary of the BH shadow. On substituting $\gamma = \Lambda = Q = 0$, the above impact parameters in Eqs. \eqref{BS14} and \eqref{BS15}, exactly reduces to the case of Kerr BH
\begin{eqnarray}\label{BS16}
&&\xi=\frac{a^2 (M+r_0)+r_0^2 (r_0-3 M)}{a (M-r_0)},\\
&&\eta=-\frac{r_0^3 \left(r_0 (r_0-3 M)^2-4 a^2 M\right)}{a^2 (M-r_0)^2}.
\end{eqnarray}
To define shadow on a celestial sphere for an observer at special infinity, we define the celestial coordinates of shadow and will find out the corresponding contours of the BH.
\section{Shadow of KNdSQ BH}    
\label{sec:shadow}                   
This section is devoted to the study of BH shadow. Since an observer at a special infinity can distinguish the shadow of a BH over a bright background, as a dark region. In principle, the shadow of stationary BHs occurs just like a circular disc, whereas spin parameter causes distortion to the shadow of spinning BHs. The apparent shape of a BH image can be preferably visualized with the help of celestial coordinates $\alpha$ and $\beta$, which could be defined as \cite{Hioki,Chandrasekhar} 
\begin{eqnarray}\label{BS17}
&&{\alpha}= \lim_{r_{\star} \to \infty} \left(-r_\star^2 \sin\vartheta \frac{d\phi}{dr} \right),\\\label{BS18}
&&{\beta}= \lim_{r_{\star} \to \infty}\left(r_\star^2 \frac{d\phi}{dr} \right).
\end{eqnarray}
In the above expressions, $r_{\star}$ represents the distance between BH and observer, while $\vartheta$ denotes the angle of inclination between the observer’s line of sight and BH spinning axis. By making use of the geodesics equations, the above Eqs. \eqref{BS17} and \eqref{BS18}, simplifies to
\begin{eqnarray}\label{BS19}
&&{\alpha}= -{\xi}{\csc \vartheta},\\\label{BS20}
&&{\beta}=\pm \sqrt{\eta+a^2\cos^2\vartheta-\xi^2 \cot^2\vartheta}.
\end{eqnarray}
Since the shadow of a BH is remarkable on the equatorial plane, therefore by substituting $\vartheta=\pi/2$, we obtain
\begin{eqnarray}\label{BS}
&&{\alpha}=-\xi,\\
&&{\beta}=\pm \sqrt{\eta}
\end{eqnarray}
\begin{figure*}
\begin{minipage}[b]{0.55\textwidth} \hspace{0.3cm}
        \includegraphics[width=0.7\textwidth]{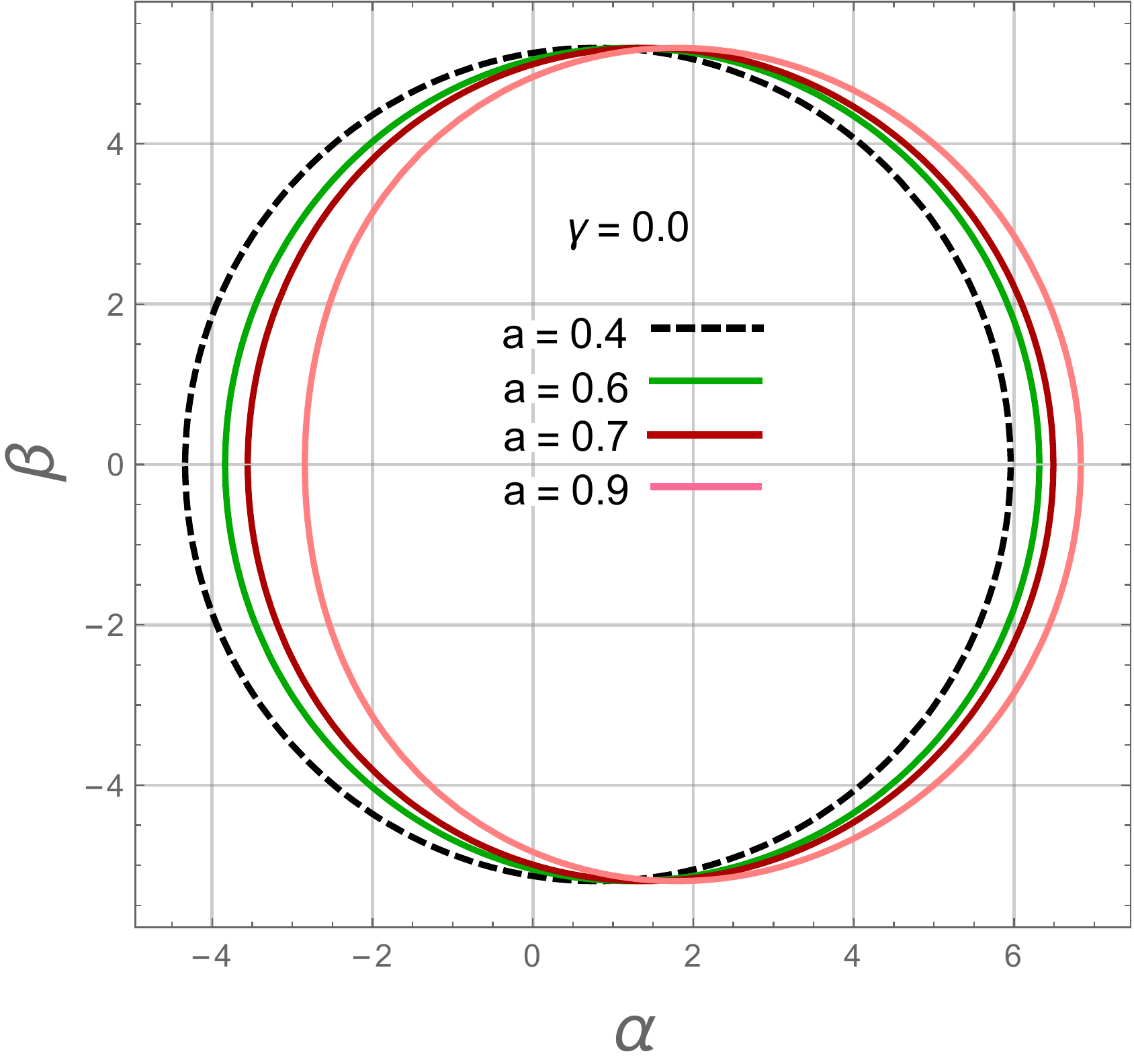}
    \end{minipage}
    \vspace{0.3cm}
        \begin{minipage}[b]{0.55\textwidth} \hspace{-1.4cm}
       \includegraphics[width=.7\textwidth]{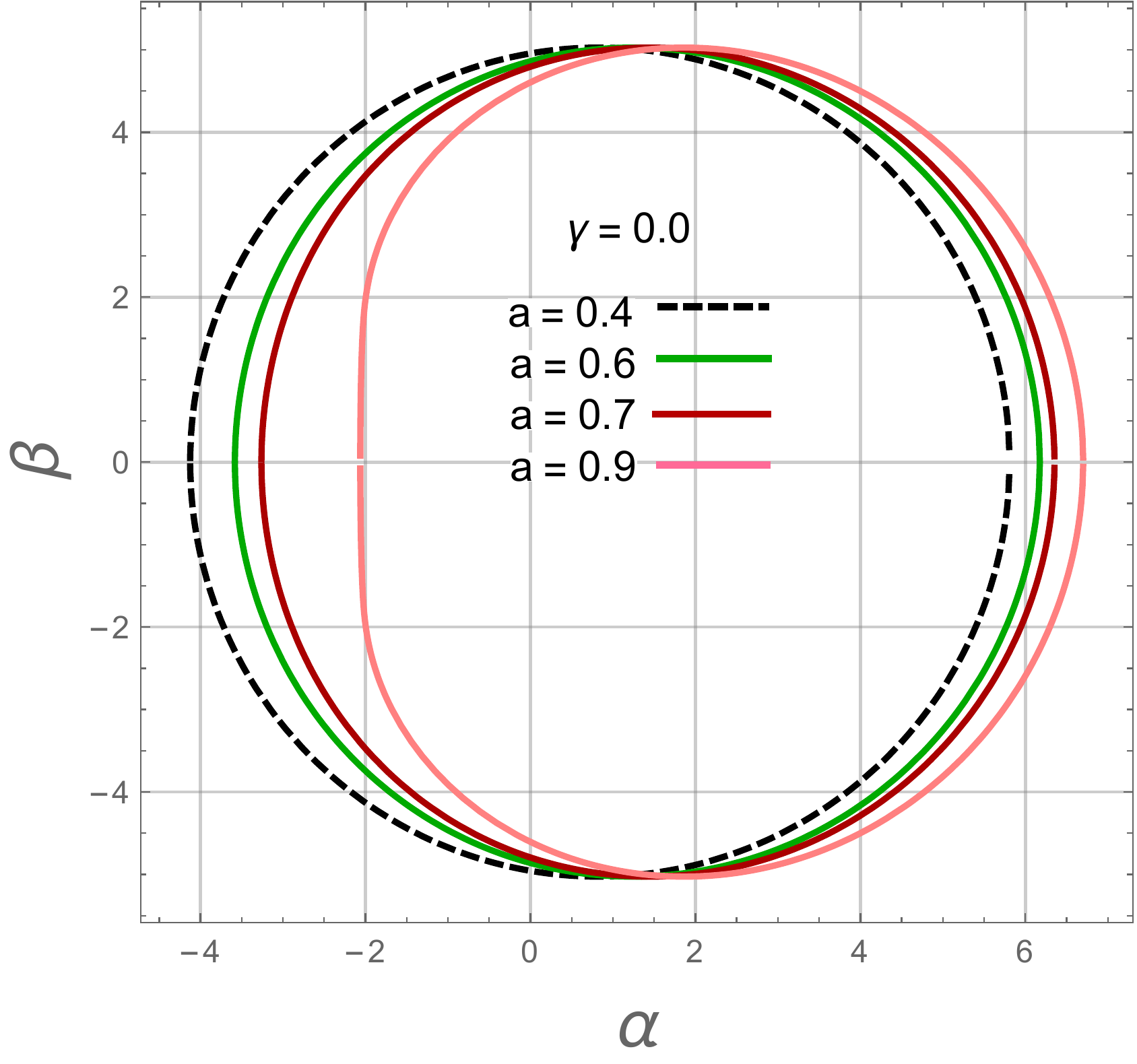}
    \end{minipage}
\begin{minipage}[b]{0.55\textwidth} \hspace{0.3cm}
        \includegraphics[width=0.7\textwidth]{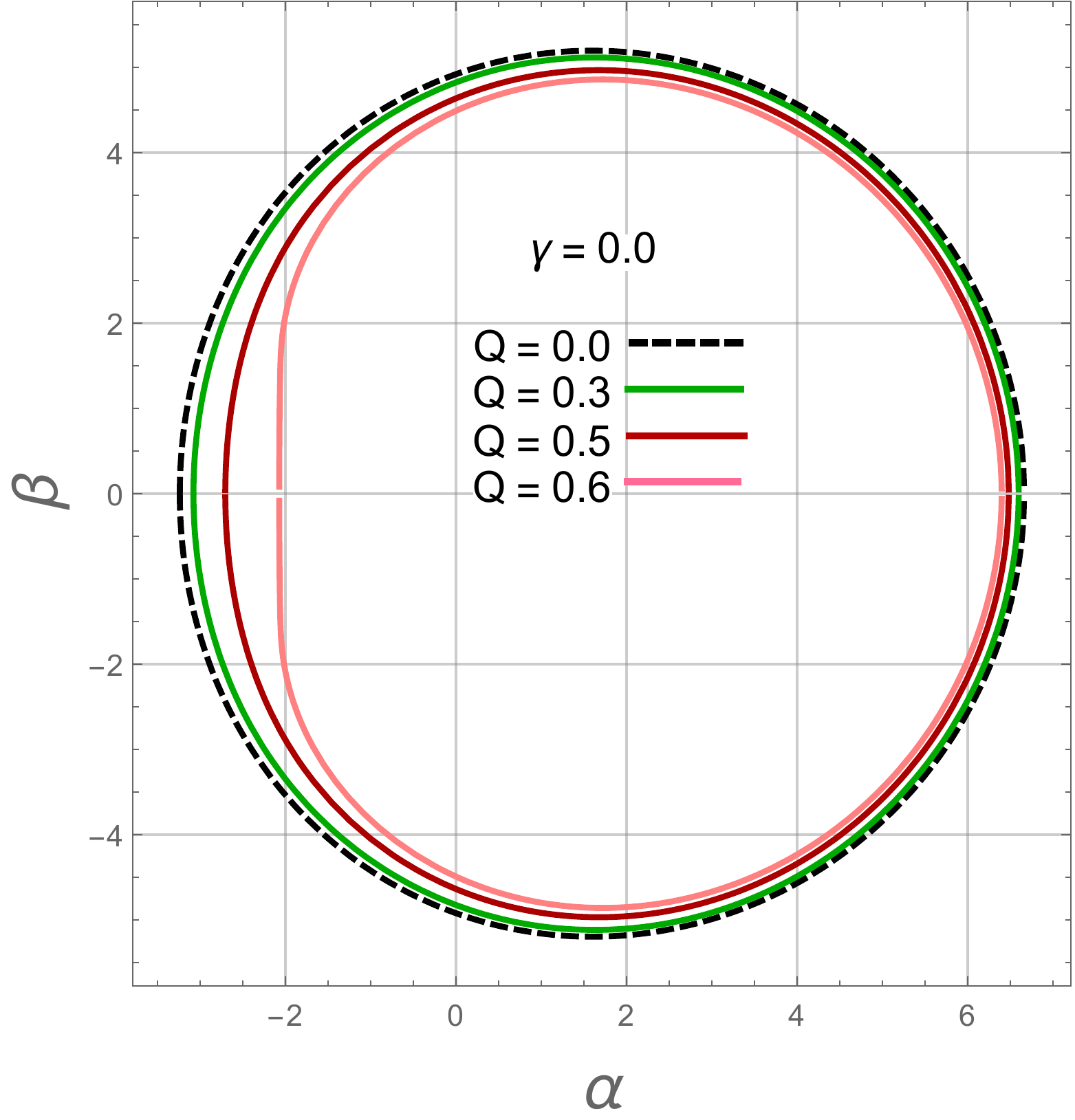}
    \end{minipage}
        \begin{minipage}[b]{0.56\textwidth} \hspace{-1.4cm}
       \includegraphics[width=.7\textwidth]{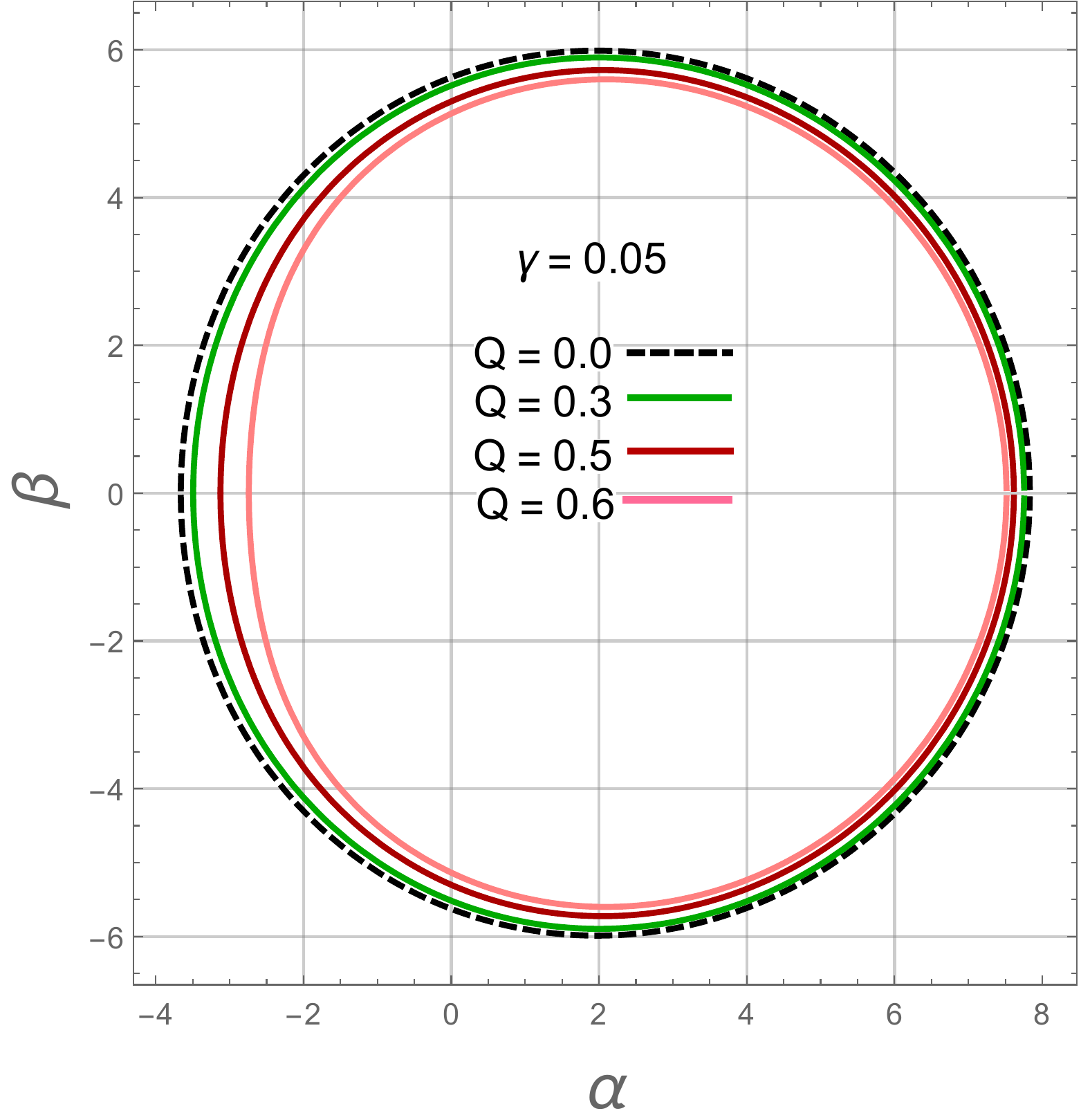}
    \end{minipage}
\caption{The upper row is plotted for $Q=0$ (left panel) and for $Q=0.435$ (right panel), while the lower row is plotted for $a=0.8$ both left, as well as right panels.}\label{Shadow1}
\end{figure*}
\begin{figure*}
\begin{minipage}[b]{0.55\textwidth} \hspace{0.3cm}
        \includegraphics[width=0.7\textwidth]{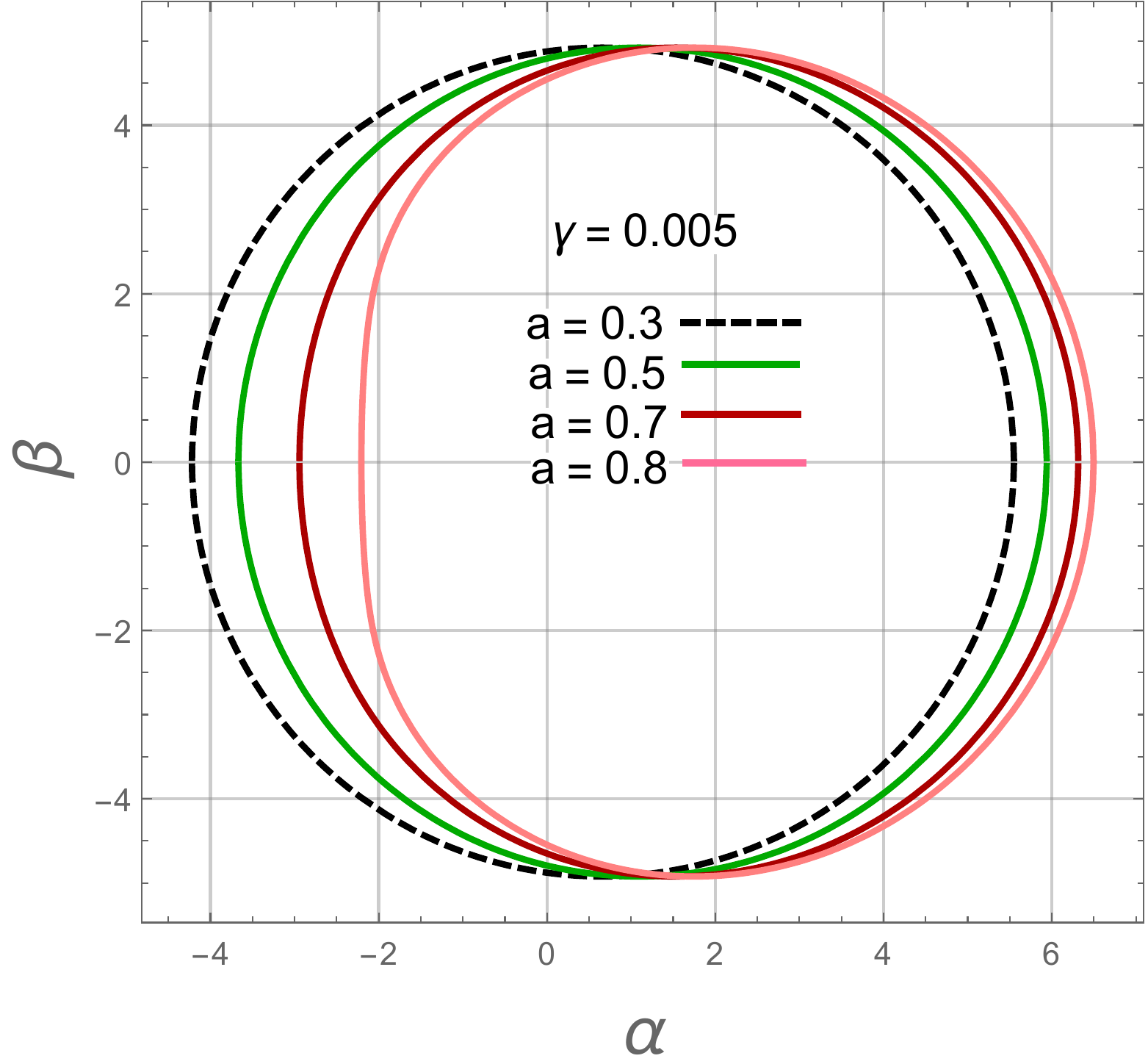}
    \end{minipage}
        \begin{minipage}[b]{0.555\textwidth} \hspace{-1.4cm}
       \includegraphics[width=.735\textwidth]{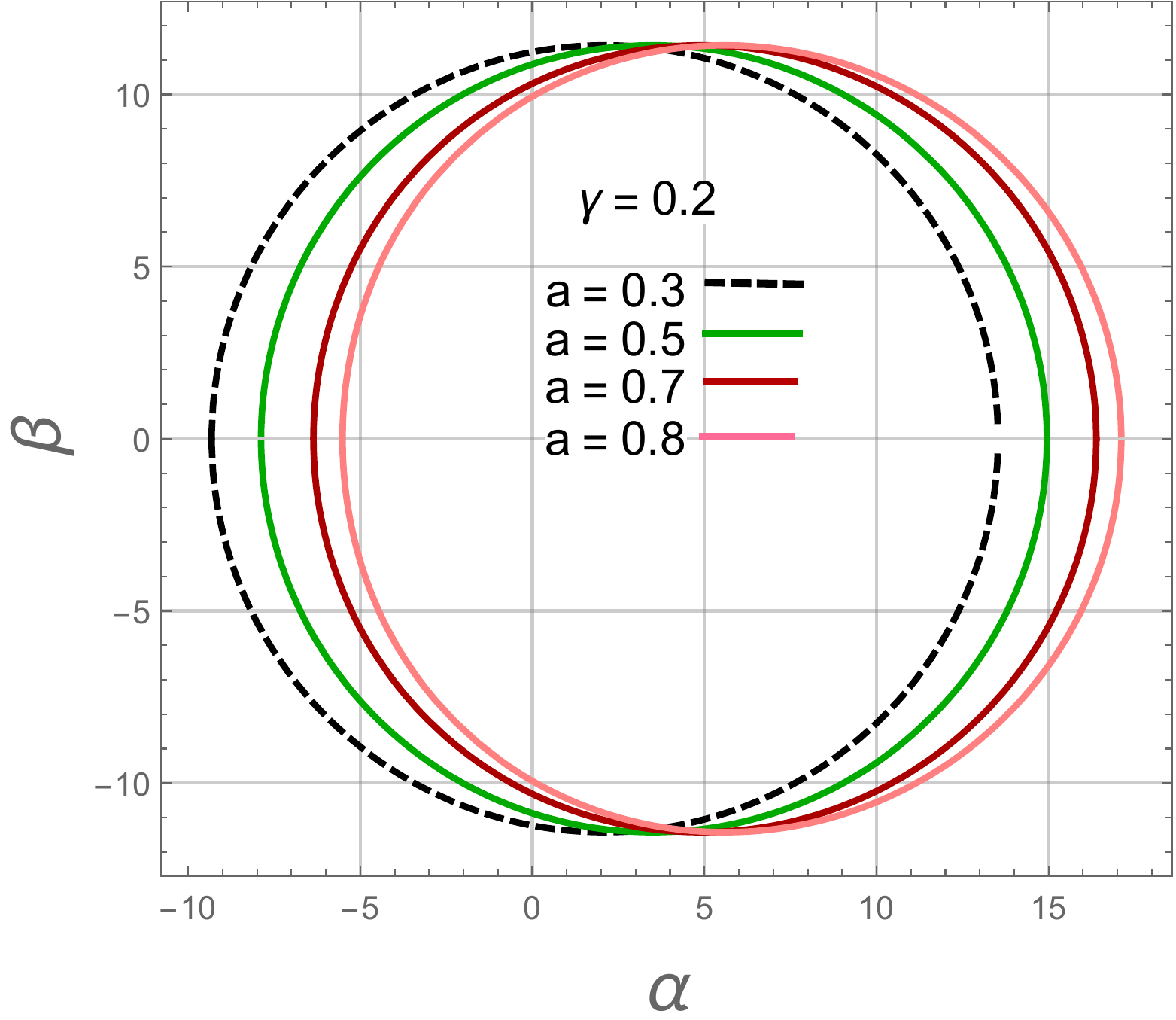}
    \end{minipage}
    \caption{Shadow cast by the KNdSQ BH at $Q=0.6$ both left, as well as right panels.}\label{Shadow2}
    \end{figure*}
In Fig. \ref{Shadow1}, the graphical description of BH shadow illustrates that due to the dragging effect, BH spin elongates its shadow in the direction of the rotation axis. Moreover, the presence of BH charge distorts its shadow, as far as it rotates faster. On the other hand, BH charge $Q$ results in decreasing the size of its image at a fixed value of the spin parameter. In Fig. \ref{Shadow2}, again the plotted behaviour shows that BH spin elongates its shadow towards the right, at different values of the quintessence field $\gamma$. From there, we discovered that the smaller value of $\gamma$ does not have enough effect on the shadow, but higher values of $\gamma$ have much affect on its behaviour. At $\gamma=0.2$, the elongation becomes larger and the shadow appears like a circular disc. 
\begin{figure*}
\begin{minipage}[b]{0.55\textwidth} \hspace{0.3cm}
        \includegraphics[width=0.7\textwidth]{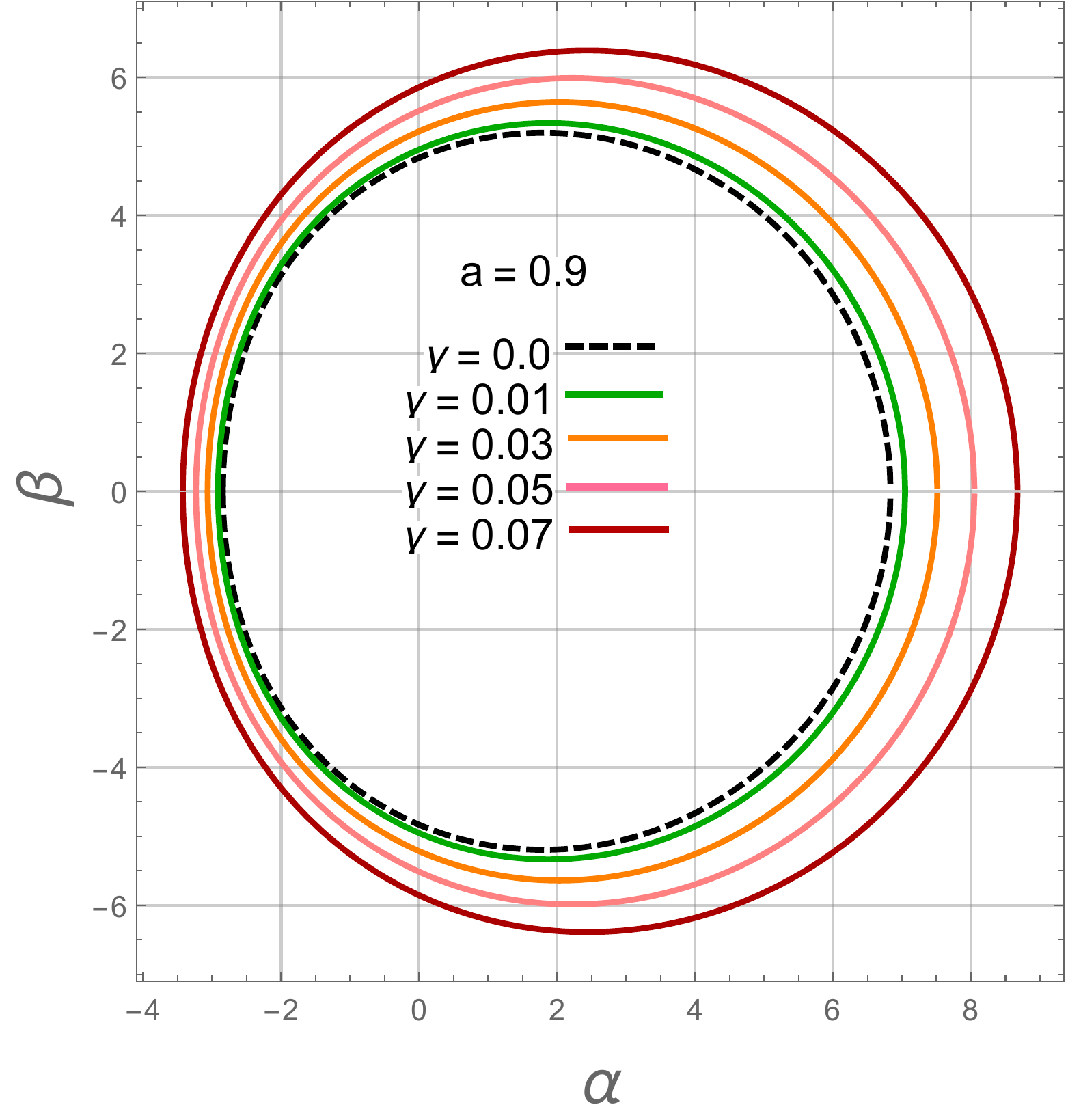}
    \end{minipage}
    \vspace{0.3cm}
        \begin{minipage}[b]{0.55\textwidth} \hspace{-1.4cm}
       \includegraphics[width=0.7\textwidth]{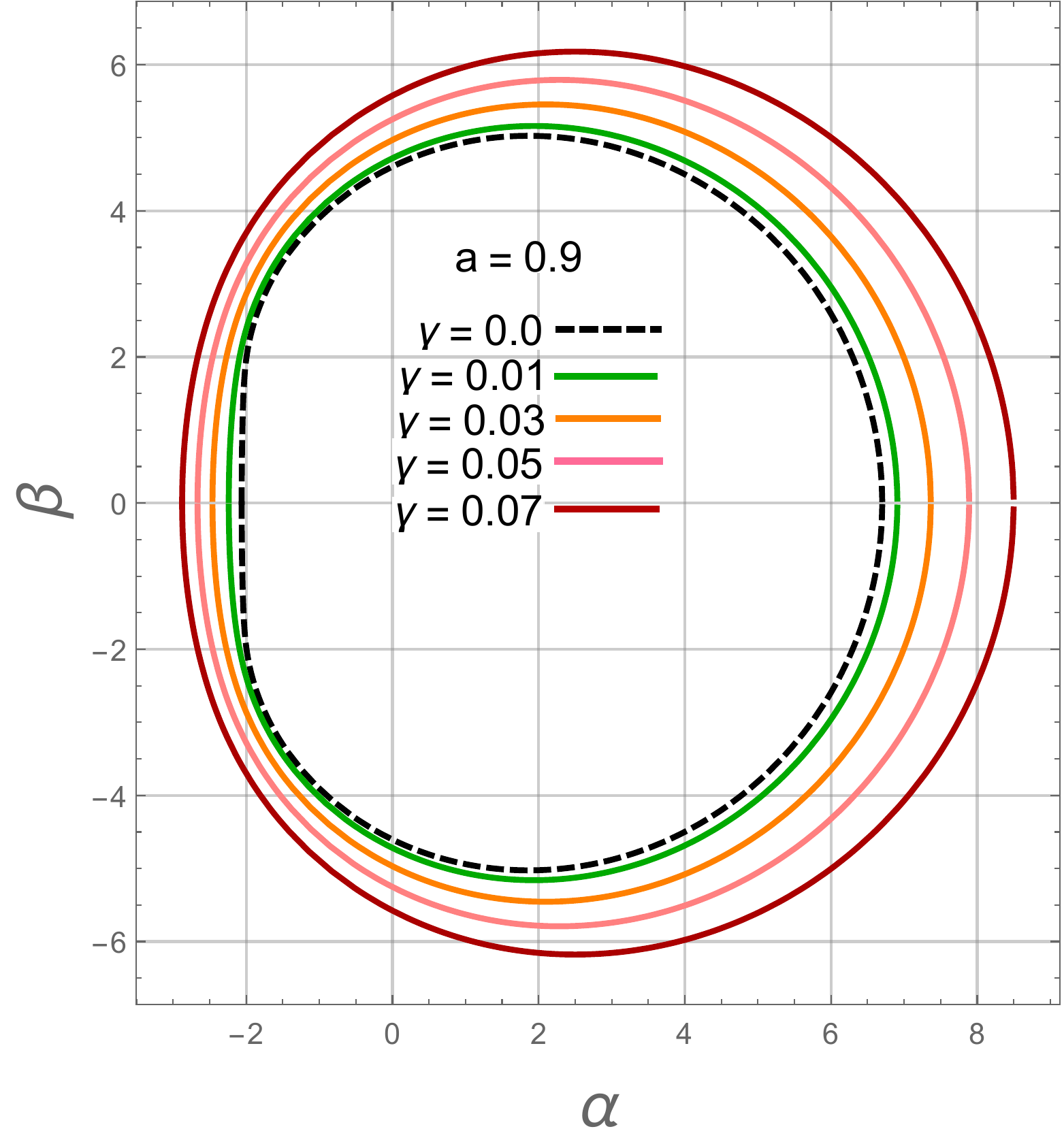}
    \end{minipage}
\begin{minipage}[b]{0.55\textwidth} \hspace{0.3cm}
        \includegraphics[width=0.7\textwidth]{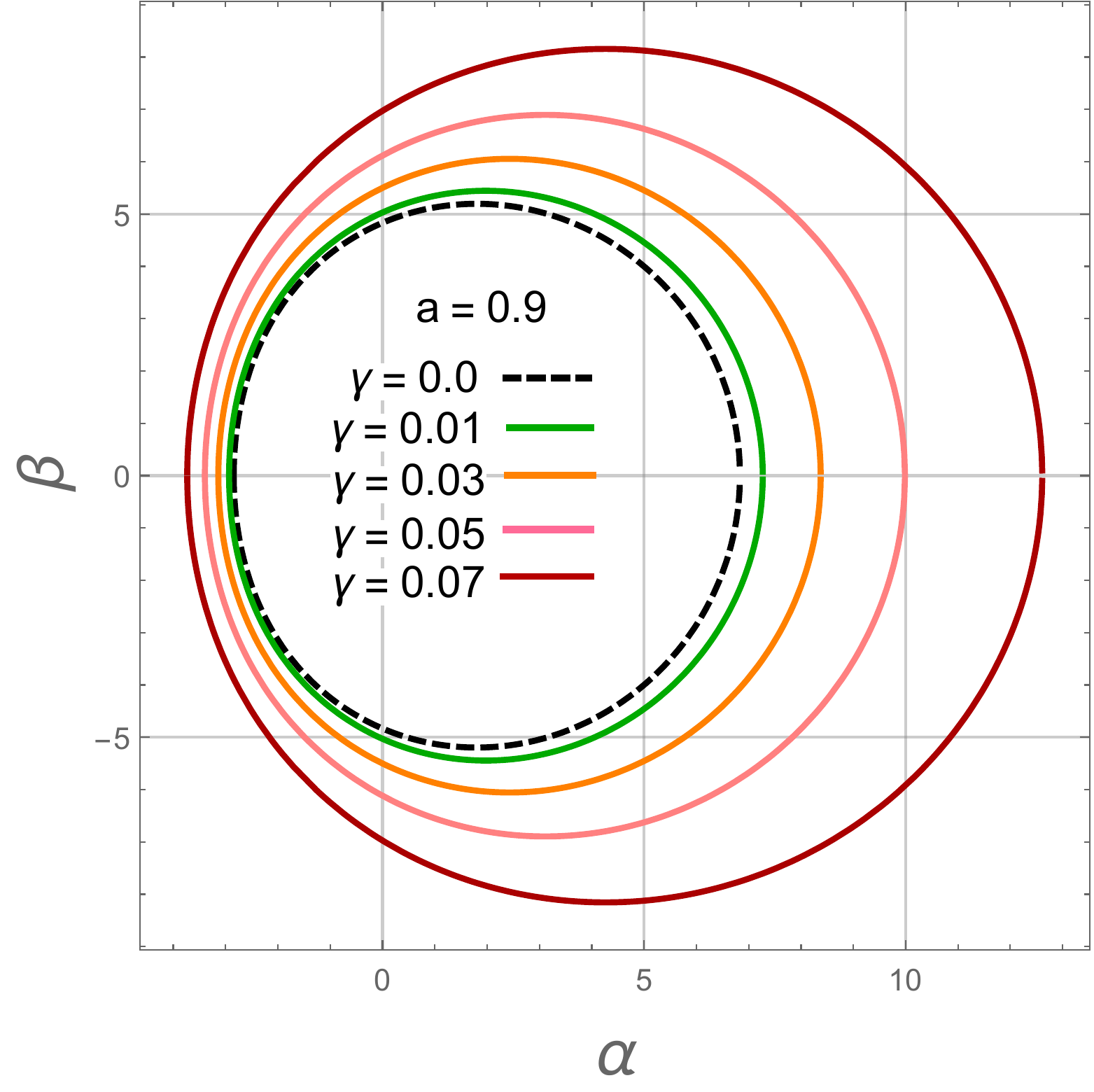}
    \end{minipage}
        \begin{minipage}[b]{0.55\textwidth} \hspace{-1.4cm}
       \includegraphics[width=0.69\textwidth]{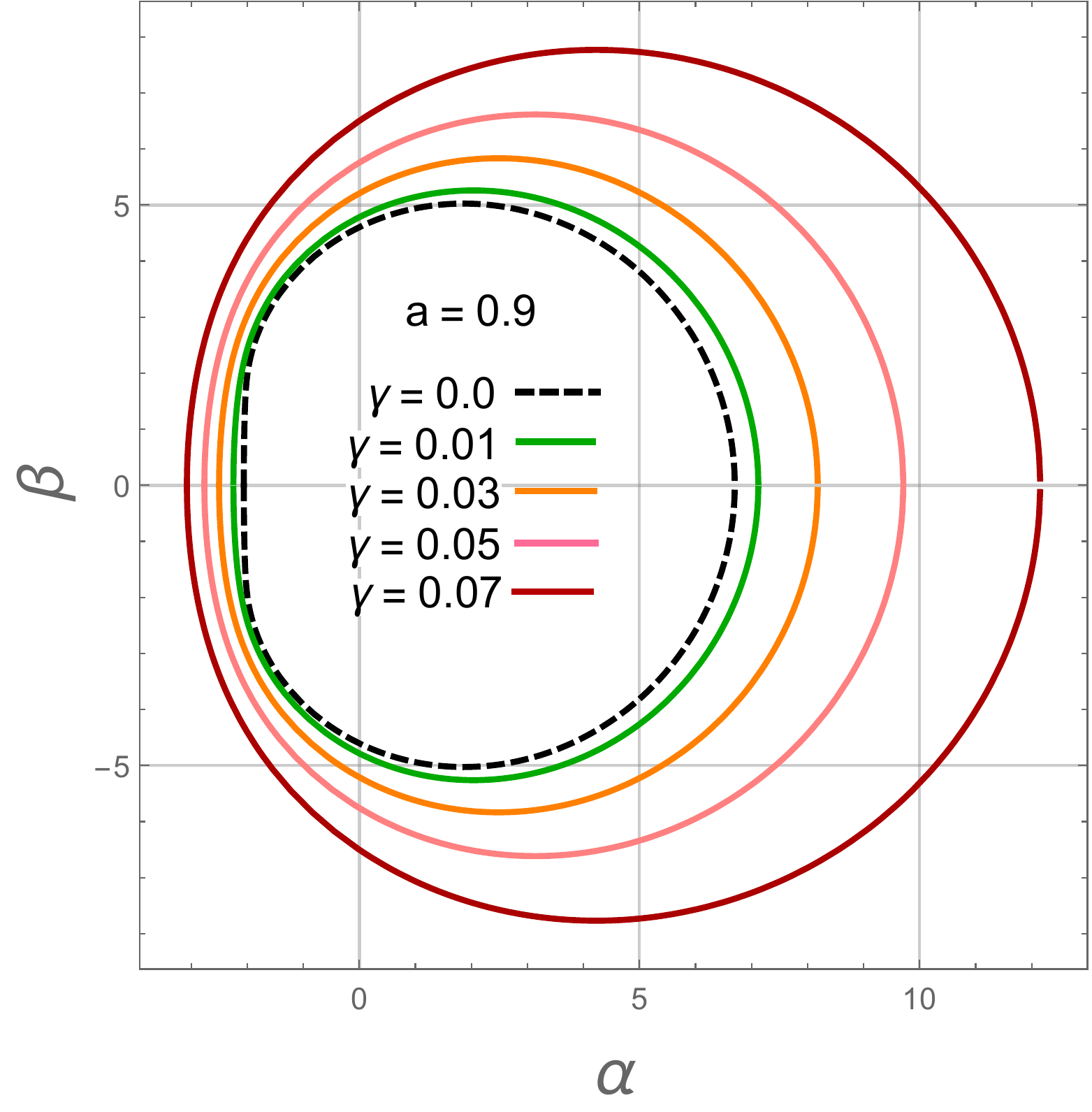}
    \end{minipage}
    \caption{The upper and bottom rows are plotted at $\omega=-1/2$ and $\omega=-2/3$, respectively. For $Q=0$ in the left panels, while $Q=0.435$ in the right panels.}\label{Shadow3}
    \end{figure*}
Figure \ref{Shadow3} reflects that the presence of BH charge leads to a more distorted shadow, as compared to the case of $Q=0$. Furthermore, at a fixed value of the BH spin $a$, the intensity of the quintessence field enlarges the size of BH shadow. 

\begin{figure*}
\begin{minipage}[b]{0.58\textwidth} \hspace{0.4cm}
        \includegraphics[width=0.7\textwidth]{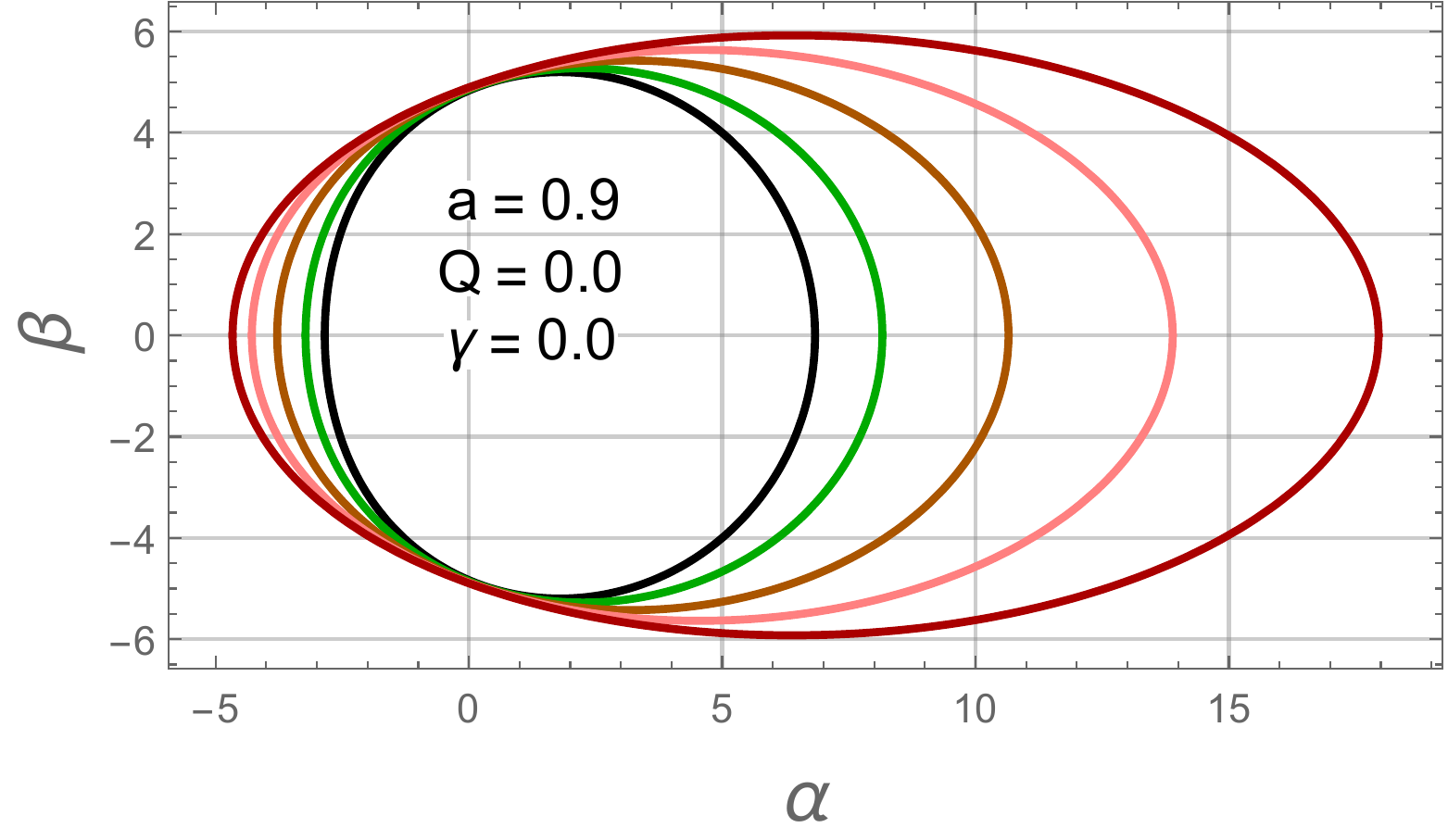}
    \end{minipage}
    \vspace{0.3cm}
        \begin{minipage}[b]{0.58\textwidth} \hspace{-1.4cm}
       \includegraphics[width=0.7\textwidth]{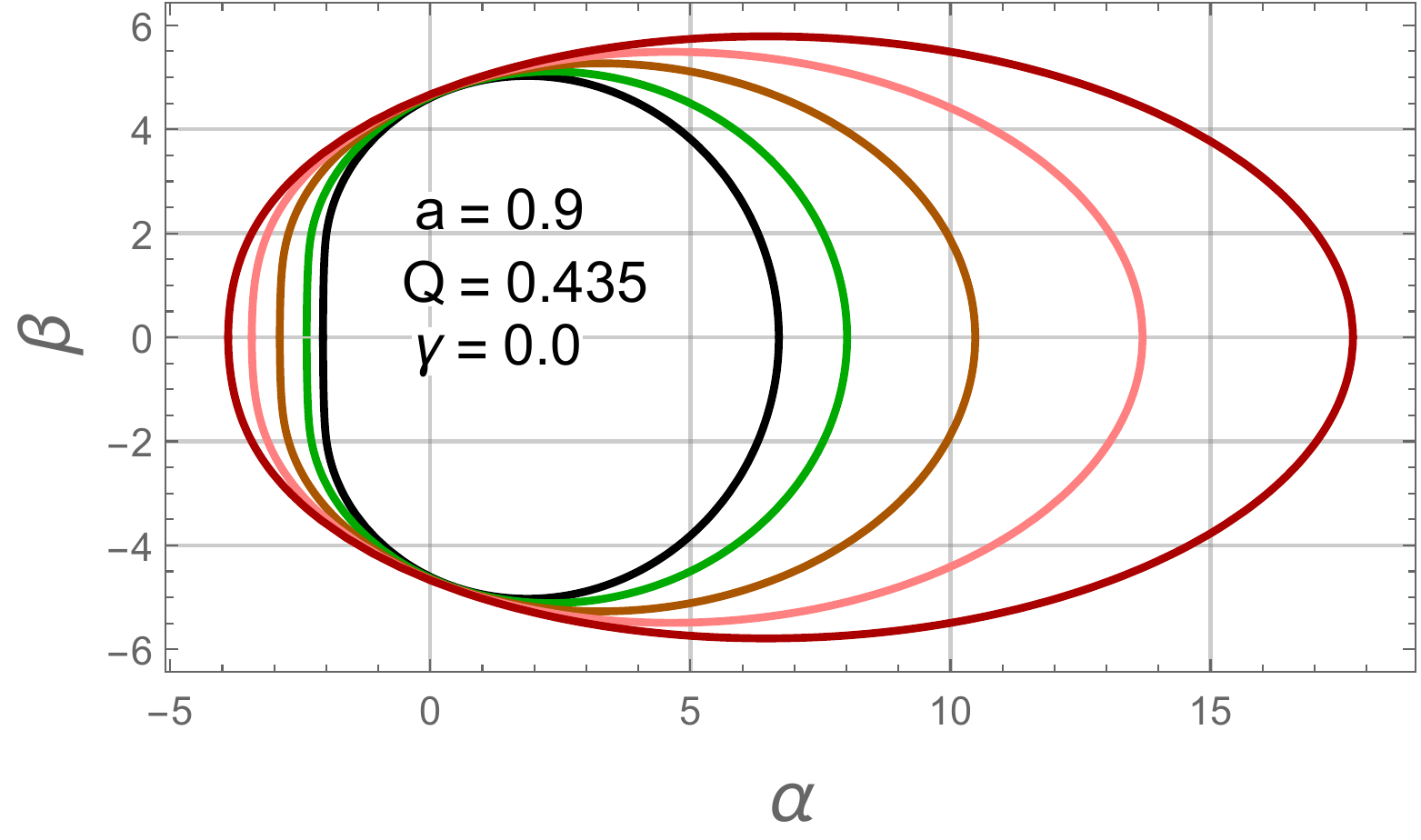}
    \end{minipage}
\begin{minipage}[b]{0.58\textwidth} \hspace{0.4cm}
        \includegraphics[width=0.7\textwidth]{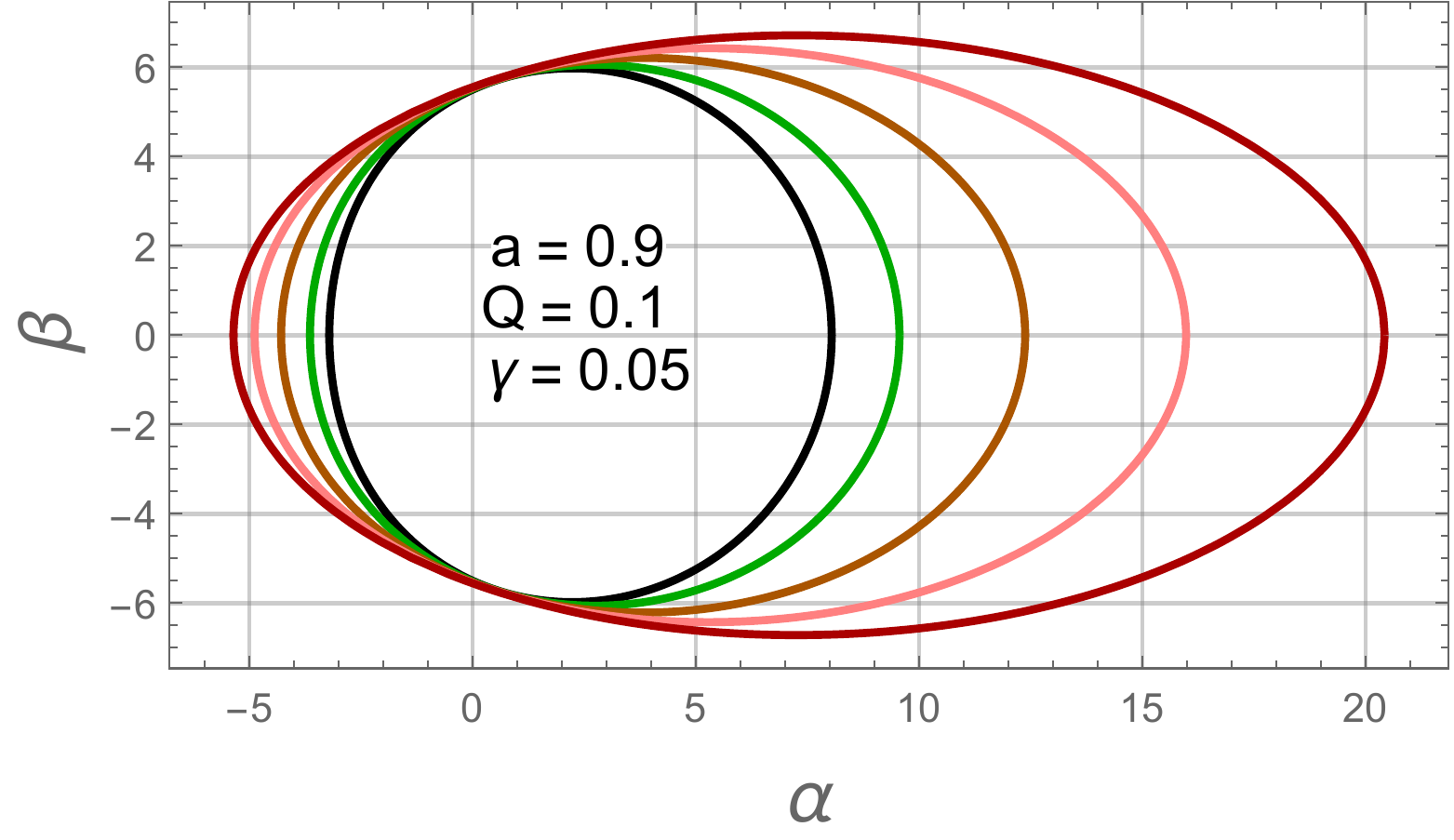}
    \end{minipage}
        \begin{minipage}[b]{0.58\textwidth} \hspace{-1.4cm}
       \includegraphics[width=0.7\textwidth]{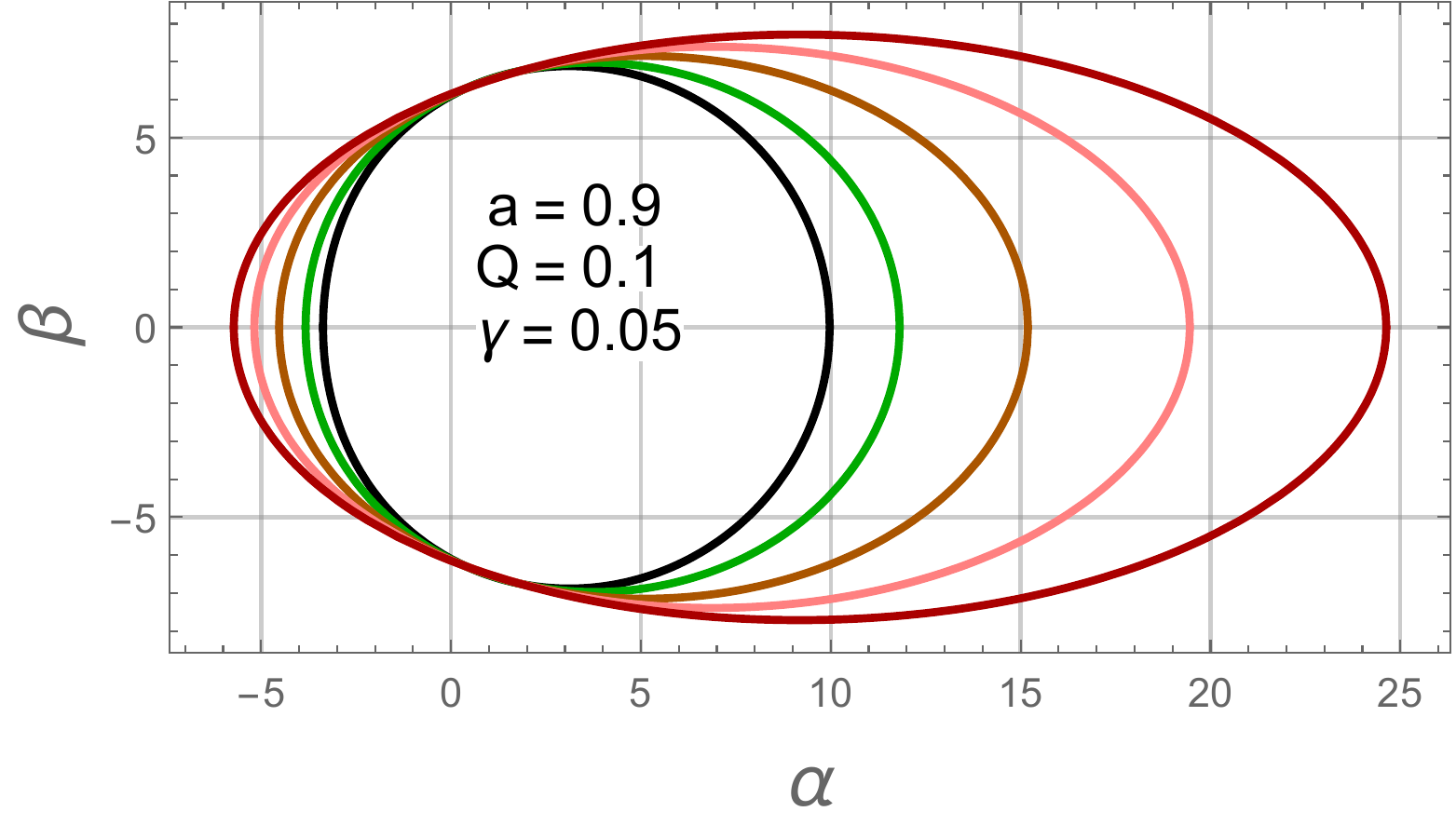}
    \end{minipage}
    \caption{All the figures are plotted at $\theta=\pi/2, \pi/3, \pi/4, \pi/5$ and $\pi/6$, respectively from inner to the outer rings. The bottom left panel is at $\omega=-1/2$ and right one at $\omega=-2/3$. }\label{Shadow4}
    \end{figure*}

Figure \ref{Shadow4} is plotted for different values of the inclination angle $\vartheta$. In the upper row, one can note that BH charges make the shadow more distorted (right panel), as compared to the chargeless case (left panel). The lower row shows that the effect of quintessence DE makes the apparent shadow wider, while its wideness also depends on the value of $\omega$.
\\
{\bf Note that}: Since the cosmological constant $\Lambda\approx 1.3\times 10^{-56} cm^{-2}$ is too small and its effect is almost negligible, hence we ignored its effect in the current study.
\subsection{Observables}
\label{subsec:observables}
\label{sec:5}
\begin{figure*}
\begin{minipage}[b]{0.58\textwidth} \hspace{0.2cm}
        \includegraphics[width=0.75\textwidth]{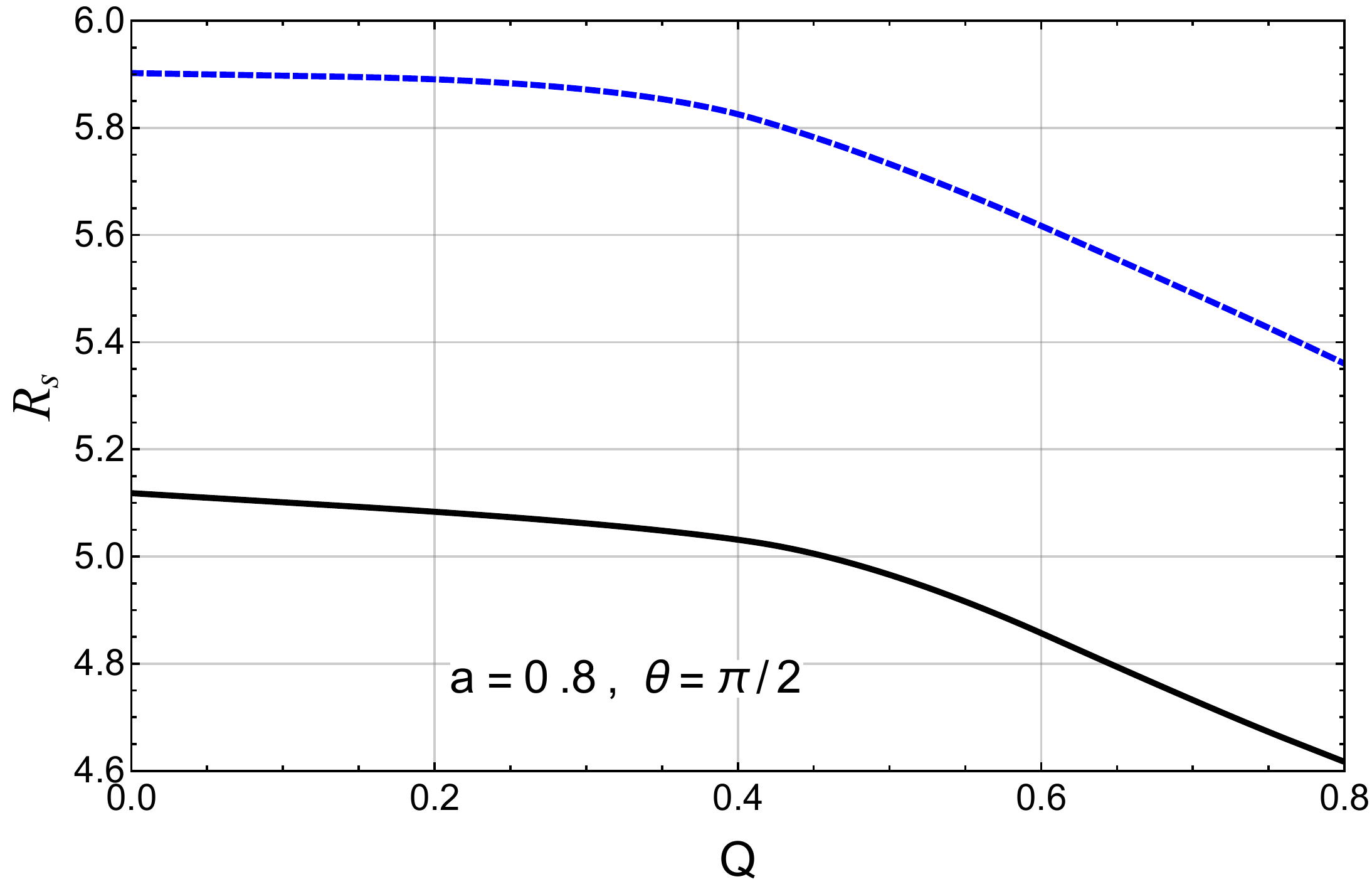}
    \end{minipage}
        \begin{minipage}[b]{0.58\textwidth} \hspace{-1.3cm}
       \includegraphics[width=0.75\textwidth]{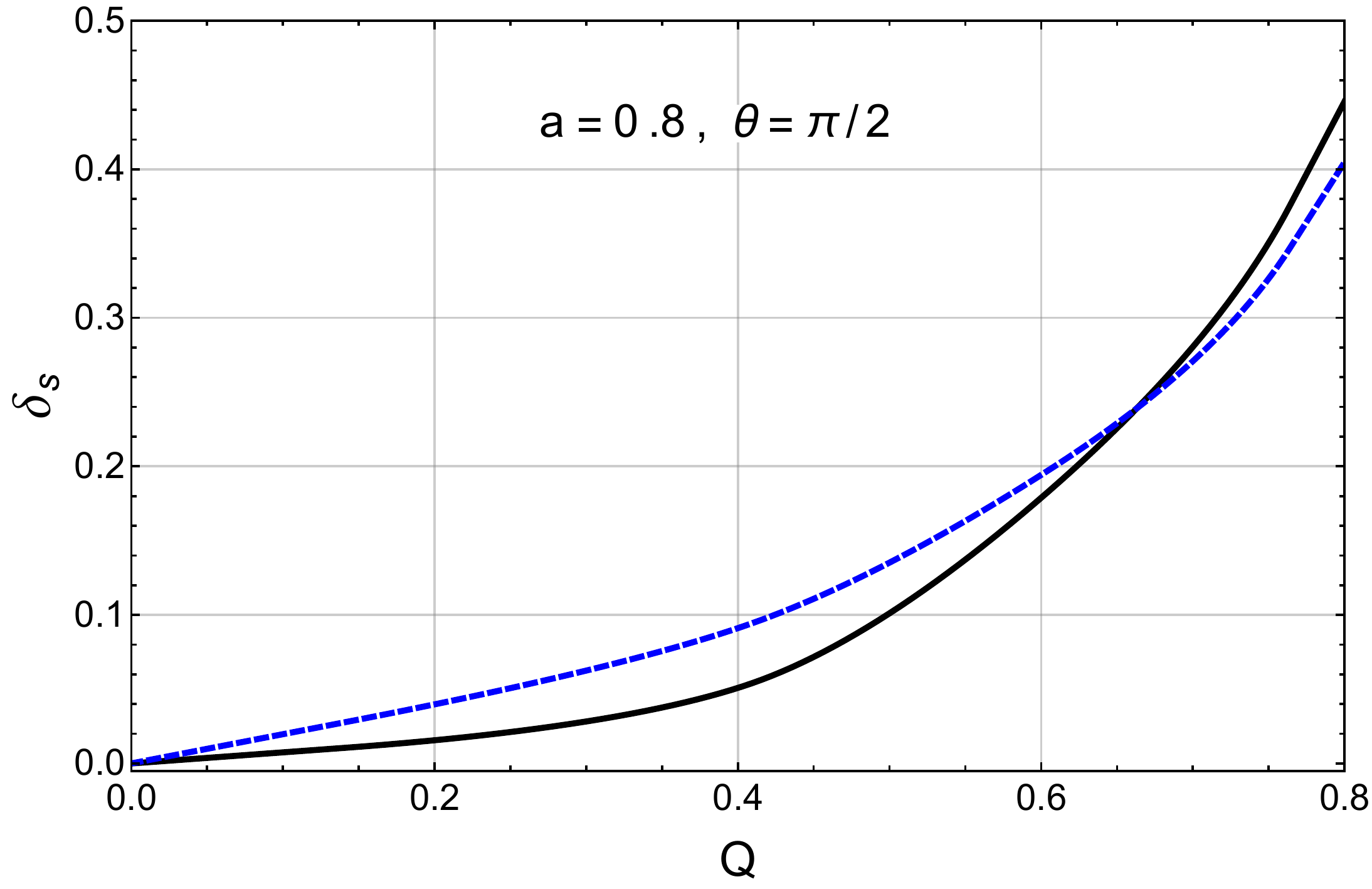}
    \end{minipage}
    \caption{Plots describing the behaviours of $R_s$ and $\delta_s$ along with BH charge $Q$, at $\gamma=0$ (black curves) and $\gamma=0.05$ (blue dashed curves).}\label{Observable}
    \end{figure*}

For illustrating the apparent shape of a rotating charged BH in quintessential DE, we outline two observables, i.e., the shadow radius $R_s$ and the distortion parameter $\delta_s$ \cite{Hioki}. Indeed, $R_s$ is approximated by the radius of a reference circle passing through three points on the shadow periphery, while $\delta_s$ define shadow's deviation from a perfect circle. Mathematically they can be approximated as
\beq
R_s=\frac{(\alpha_t-\alpha_r)^2+\beta_t^2}{2 \mid \alpha_t-\alpha_r \mid}, \quad \delta_s=\frac{\mid \tilde{\alpha_l}-\alpha_l \mid}{R_s}.
\eeq
In the above expressions, ($\alpha_t, \beta_t$) and ($\alpha_r, \beta_r$) are, respectively the topmost and rightmost points where the reference circle cut the shadow, whereas ($\tilde{\alpha_l},0$) and ($\alpha_l,0$) represents the points, at which the reference circle and shadow intersect the leftmost $\alpha$ axes. Figure \ref{Observable} describes the behaviour of observables
$R_s$ and $\delta_s$, for the KN and KNdSQ BHs. We found that BH charge $Q$ diminishing the shadow radius $R_s$, while increases the distortion parameter $\delta_s$. Moreover, we observed that the presence of quintessential DE contributes to the shadow radius.
\section{Conclusion}
\label{sec:conclusion}
This article aimed to establish the theoretical investigation of BH shadow for a rotating charged BH in quintessential DE. We hope that our investigation would be helpful for the upcoming Event Horizon Telescope observations. We have started our investigation by reviewing the spacetime geometry of a KNdSQ BH and determined the necessary expressions for the exploration of BH shadow. Firstly, we have investigated BH horizons and the deflection angle of light using the GBT, which is found to be
\begin{equation}\nonumber
\hat{\Theta} \equiv \frac{4M}{b}+\frac{2\gamma}{b}-\frac{3\gamma\omega}{b}-\frac{3Q^2}{4b^2}\pm\frac{4aM}{b^2}.
\end{equation}
We have observed that by letting $\gamma=Q=0$, the deflection angle of light reduces to the Kerr BH, while to the Schwarzschild BH if $a=\gamma=Q=0$. Secondly, by utilizing the Hamiltonian-Jacobi approach, we have derived the null geodesics. Moreover, we have introduced celestial coordinates to construct the images of BH shadow, as they are very essential for the construction of BHs apparent shadow. For the graphical demonstration, we have plotted different cases of the BH shadow.

Apart from BH charge and spin parameters, our main interest was to investigate the effect of quintessential DE on the shadow cast by a KNdSQ BH. The effect of quintessential DE around a BH has evaluated on its shadow in detail. The obtained result demonstrates that due to the dragging effect, BH spin elongates its shadow in the direction of the rotational axis, while increases the deflection angle of light. BH charge reduces both of its shadow radius and angle of deflection. Moreover, both $a$ and $Q$ significantly increases the distortion effect in BH shadow. Geometrically the shadow size increases with the increase of $\gamma$, while at $\gamma= 0.2$ it appeared like a perfect circular disc. As shown in Fig. \ref{Shadow4}, the shadow of KNdSQ BH horizontally elongates as the angle of inclination decreases.
\subsubsection*{Acknowledgment}
This work is supported by the National Natural Science Foundation of China (11771407).

\bibliographystyle{unsrt}

\end{document}